\documentclass[amsmath,eqsecnum,preprint,pre,aps,showpacs]{revtex4}
\usepackage{amsmath}
\usepackage{graphicx}
\usepackage{epstopdf}
\usepackage{dcolumn}
\usepackage{bm}
\usepackage{float}

\begin{document}

\title{Colloidal liquids of yolk-shell particles}
\author{L. E. Sanchez Diaz{$^\dagger$}, E. C.
Cortes-Morales{$^\ddagger$}, X. Li{$^\dagger$},Wei-Ren Chen{$^\dagger$}, and M. Medina-Noyola{$^\ddagger$} }

\address{$^\dagger$Biology and Soft Matter Division, Oak Ridge
National Laboratory, Oak Ridge, Tennessee 37831, USA}
\address{$^\ddagger$Instituto de F\'{\i}sica {\sl ``Manuel Sandoval
Vallarta"}, Universidad Aut\'{o}noma de San Luis Potos\'{\i},
\'{A}lvaro Obreg\'{o}n 64, 78000 San Luis Potos\'{\i}, SLP,
M\'{e}xico}
\date{\today}

\begin{abstract}

In this paper we develop statistical mechanical tools to describe
the intermediate- and long-time collective- and self-diffusion
properties of a liquid of strongly-interacting hollow spherical
particles (shells), each bearing a smaller solid sphere (yolk) in
its interior. To decouple two complex effects we assume that the
hydrodynamic interactions can be accounted for through the
effective short-time self-diffusion coefficients $D^0_s$ and
$D^0_y$  that describe the short-time Brownian motion of the yolk
and the shell particles, and develop a self-consistent generalized
Langevin equation theory to describe the intermediate- and
long-time effects of the direct shell-shell, yolk-shell and
yolk-yolk interactions. In a concrete application, we consider the
simplest yolk-shell model system involving purely repulsive
hard-body interactions between all (shell and yolk) particles.
Using a softened version of these interparticle potentials we
perform Brownian dynamics simulations to determine the mean
squared displacement of both types of particles, as well as the
intermediate scattering function of the yolk-shell complex. We
compare the theoretical and simulation results between them, and
with the results for the same system in the absence of yolks. We
find that the yolks, which have no effect on the shell-shell
static structure, influences the dynamic properties in a
predictable manner, fully captured by the theory.

\end{abstract}


\maketitle

\section{Introduction}

The increased ability to manufacture colloidal particles with
specific morphologies has presented colloid science with many new
and fascinating challenges and opportunities. One example is
provided by yolk-shell particles, i.e., hollow spherical particles
(shells) bearing a smaller solid sphere (yolk) in its interior
\cite{lou,kam}. Sophisticated nanostructures with this morphology
can be synthesized for application in various fields such as
biomedical imaging \cite{liu}, catalysis \cite{li}, and energy
storage devices \cite{liu}. Understanding the diffusive properties
of individual yolk-shell particles when they operate in aqueous
environments represents an interesting challenge. Manipulating
their eventual self-assembly \cite{langmuirmagnetic}, or simply
considering realistic conditions, requires a god understanding of
the structure and dynamics of systems of interacting yolk-shell
particles in terms of the yolk-yolk, yolk-shell, and shell-shell
direct (i.e., conservative) and hydrodynamic interactions. It is
not difficult to envision possible applications in many areas of
materials design and engineering, of the condensed phases
(liquids, crystals, and even glasses) that could be formed with
yolk-shell particles.

One then would like to describe the effects of these interactions
on, for example, the individual and the collective diffusive
properties of yolk-shell particles immersed in a solvent at finite
concentration. The dynamic properties of ordinary colloidal
dispersion refer to the relaxation of the fluctuations $\delta
n({\bf r} ,t)$ of the local concentration $n({\bf r},t)$ of
colloidal particles around its bulk equilibrium value $n$. The
average decay of $\delta n({\bf r},t)$ is described by the
time-dependent correlation function $F(k,t)\equiv \left\langle
\delta n({\bf k},t)\delta n(-{\bf k},0)\right\rangle $ of the
Fourier transform $\delta n({\bf k},t)\equiv (1/N)\sum_{i=1}^N
\exp {[i{\bf k}\cdot{\bf r}_i(t)]}$ of the fluctuations $\delta
n({\bf r} ,t)$, with ${\bf r}_i(t)$ being the position of particle
$i$ at time $t$. $F(k,t)$ is referred to as the intermediate
scattering function, measured by experimental techniques such as
dynamic light scattering \cite{pusey,brader}. One can also define
the mean squared displacement $W(t)\equiv <(\Delta{\bf
R}(t))^2>/6$ and the self-intermediate scattering function,
$F_S(k,t)\equiv \left\langle \exp {[i{\bf k}\cdot \Delta{\bf
R}(t)]} \right\rangle $, where $\Delta{\bf R}(t)$ is the
displacement of any of the $N$ particles over a time $t$. The
experimental and theoretical study of the dynamics of ordinary
colloidal dispersion can be said to be fairly well established
\cite{pusey,brader}. One then would like to extend now the same
level of understanding to yolk-shell suspensions. A basic and
elementary question, for example, refers to the effect of the
yolk-shell interaction on the Brownian motion of the complex, as
exhibited by the difference between the Brownian motion of an
empty shell, and of a shell carrying its yolk. A second question
refers to the effects of (direct and hydrodynamic) interactions
between yolk-shells on the individual and collective diffusion of
a concentarted suspension of strongly-interacting yolk-shell
particles.

The detailed and simultaneous description of the coupled effects
of direct and hydrodynamic interactions is, in general, a highly
involved problem. We may profit, however, from the ingenious
manners to decouple them, developed in similar challenges in
colloid science \cite{pusey,brader}. For example, at least for
rigid shells, the complex effects of shell-shell hydrodynamic
interactions can be taken into account through an effective
self-diffusion coefficient $D^0_s$ describing the shells'
short-time Brownian motion \cite{prlhi,brady}. Similarly, when
focusing on the dynamics of condensed phases of
strongly-interacting yolk-shell particles, the rich phenomenology
of the hydrodynamic interactions of the captive yolk with its
confining shell could also be modelled as a simple effective
free-diffusion process, characterized by another effective
(short-time) self-diffusion coefficient $D^0_y$, also amenable to
independent experimental determination \cite{cervantes}.

In an attempt to develop the simplest model representation of this
class of materials, let us then assume that the hydrodynamic
interactions can be accounted for through the effective short-time
self-diffusion coefficients $D^0_s$ and $D^0_y$. In this context,
the main purpose of the present  work is to develop a statistical
mechanical theory of the dependence of the collective- and
self-diffusion properties of the yolk-shell suspensions, on
control parameters such as the relative mobility of the shell and
yolk particles (i.e., the ratio $D^0_s/D^0_y$), the specific
geometry  (size and thickness of the shell and size of the yolk),
and the concentration of yolk-shell particles in the suspension.
For this we develop a first-principles theory to predict the main
features of the  dynamic properties in terms of the specific
effective interactions between the shell and the  yolk particles
that constitute the system. We approach this task in the framework
of the self-consistent generalized Langevin equation (SCGLE)
theory of colloid dynamics, which is adapted here to the context
of a suspension of Brownian yolk-shell particles.

This paper is organized as follows. In Sec. II we define the model
systems considered in our study and provide the basic information
on the simulation methods employed. In Section III we adapt our
self-consistent generalized Langevin equation (SCGLE) theory to
the description of system of yolk-shell particles, and discuss
very briefly the physical content and the rationale of each of the
approximations involved in its formulation.  In Sec IV we present
and discuss the main results using Brownian dynamics simulations
and compare with the numerical solution of the full
self-consistent theory. In Sec. V we develop the SCGLE theory of
the yolk-shell systems in which yolk and shell particles are
treated on the same footing. Finally, in Sec. VI we summarize our
main conclusions.

\section{Model system and Brownian dynamics simulations}

As our fundamental starting point, let us consider a model
monodisperse yolk-shell colloidal suspension formed by $N$
spherical shell particles in a volume $V$, each of which bears one
smaller (``yolk'') particle diffusing in its interior. Let us
neglect hydrodynamic interactions and denote by ${\bf x}_{i}(t)$
and ${\bf v}_{i}(t)$ the position and velocity of the center of
mass of the $i$th shell particle (with $i=1,2,\ldots ,N$), and by
${\bf y}_{i}(t)$ and ${\bf w}_{i}(t)$ the position and velocity of
the center of the $i$th yolk particle. The microscopic dynamics of
these $2N$ Brownian particles is then described by the following
$2N$ Langevin equations,
\begin{equation}
M_s\frac{d{\bf v}_{i}(t)}{dt}= -\zeta^0_s{\bf v}_{i}(t)+{\bf
f}^{0} _{i}(t)-\sum_{j\neq i}\nabla_i u_{ss}(|{\bf x}_{i}(t)-{\bf
x}_{j}(t)|) - \sum_{j}\nabla_i u_{sy}(|{\bf x}_{i}(t)-{\bf
y}_{j}(t)|), \label{eq1}
\end{equation}
\begin{equation}
M_y\frac{d{\bf w}_{i}(t)}{dt}= -\zeta^0_y{\bf w}_{i}(t)+{\bf
g}^0_i(t)-\sum_{j\neq i}\nabla_i u_{yy}(|{\bf y}_{i}(t)-{\bf
y}_{j}(t)|) - \sum_{j}\nabla_i u_{ys}(|{\bf y}_{i}(t)-{\bf
x}_{j}(t)|), \label{eq1y}
\end{equation}
with $i=1,2,\ldots ,N$. In these equations ${\bf f}^{0}_{i}(t)$
and ${\bf g}^{0}_{i}(t)$ are Gaussian white random forces of zero
mean, and variance given by  $ \langle {\bf f}^0_{i}(t){\bf
g}^0_{j}(0)\rangle =0$,  $\langle {\bf f}^0_{i}(t){\bf
f}^0_{j}(0)\rangle =k_{B}T\zeta^0_s2\delta (t)\delta
_{ij}\stackrel{\leftrightarrow }{{\bf I}}$, and $ \langle {\bf
g}^0_{i}(t){\bf g}^0_{j}(0)\rangle =k_{B}T\zeta^0_y2\delta
(t)\delta _{ij}\stackrel{\leftrightarrow }{{\bf I}}$, with
$i,j=1,2,\ldots ,N$ and with $\stackrel{\leftrightarrow }{{\bf
I}}\mbox{being the }3\times 3 \mbox{ unit tensor}$. The radially
symmetric pairwise potentials $u_{ss}(r)$, $u_{sy}(r) =u_{ys}(r)$,
and $u_{yy}(r)$ describe, respectively, the shell-shell,
shell-yolk, and yolk-yolk direct interactions.

The microscopic dynamics represented by the $2N$ Langevin
equations in Eqs. (\ref{eq1}) and (\ref{eq1y}) constitute the
starting point of both approaches employed to describe the self
and collective dynamics of our system. Thus, the Brownian dynamics
simulations below consist essentially of the numerical solution of
these $2N$ stochastic Langevin equations. More precisely,
according to the algorithm proposed by Ermak and McCammon
\cite{ermak}, we solve the overdamped version of these equations,
in which one neglects the inertial terms $M_s[d{\bf v}_{i}(t)/dt]$
and $M_s[d{\bf w}_{i}(t)/dt]$ on the left side of Eqs. (\ref{eq1})
and (\ref{eq1y}). The resulting equations can be written as the
following prescription to generate the new positions ${\bf
x}_{i}(t+\Delta t)$ and  ${\bf y}_{i}(t+\Delta t)$ from the
current positions  ${\bf x}_{i}(t)$ and  ${\bf y}_{i}(t)$,
\begin{equation}
{\bf x}_{i}(t+\Delta t)={\bf x}_{i}(t) +\beta D^0_s {\bf
F}_{i}(t)\Delta t+ \textbf{X} _{i}(t) \label{eq1}
\end{equation}
and
\begin{equation}
{\bf y}_{i}(t+\Delta t)={\bf y}_{i}(t) +\beta D^0_y {\bf
G}_{i}(t)\Delta t+ \textbf{Y} _{i}(t), \label{eq1}
\end{equation}
with $\beta^{-1}\equiv k_BT$ being the thermal energy and $D^0_s \
(\equiv k_BT/ \zeta^0_s)$ and $D^0_y \ (\equiv k_BT/ \zeta^0_y)$
being the short-time self-diffusion coefficients of the shell and
yolk particles, respectively. In these equations ${\bf F}_{i}(t)$,
defined as
\begin{equation}
{\bf F}_{i}(t) \equiv -\sum_{j\neq i}\nabla_i u_{ss}(|{\bf
x}_{i}(t)-{\bf x}_{j}(t)|) - \sum_{j}\nabla_i u_{sy}(|{\bf
x}_{i}(t)-{\bf y}_{j}(t)|), \label{eq1}
\end{equation}
is the force exerted by all the shells and yolks  on the $i$th
shell at time $t$, and ${\bf G}_{i}(t)$, defined as
\begin{equation}
{\bf G}_{i}(t) \equiv-\sum_{j\neq i}\nabla_i u_{yy}(|{\bf
y}_{i}(t)-{\bf y}_{j}(t)|) - \sum_{j}\nabla_i u_{ys}(|{\bf
y}_{i}(t)-{\bf x}_{j}(t)|), \label{eq1y}
\end{equation}
is the force exerted by all the shells and yolks on the $i$th
yolk. The random displacements $\textbf{X} _{i}(t)\ (\equiv +\beta
D^0_s {\bf f}^{0} _{i}(t))$ and  $\textbf{Y} _{i}(t)\ (\equiv
+\beta D^0_y {\bf g}^{0} _{i}(t))$ are extracted from Gaussian
distributions with zero mean and variance given, respectively, by
$6 D^0_s \Delta t$ and $6 D^0_y \Delta t$.

\begin{figure}
\includegraphics[width=0.3\textwidth]{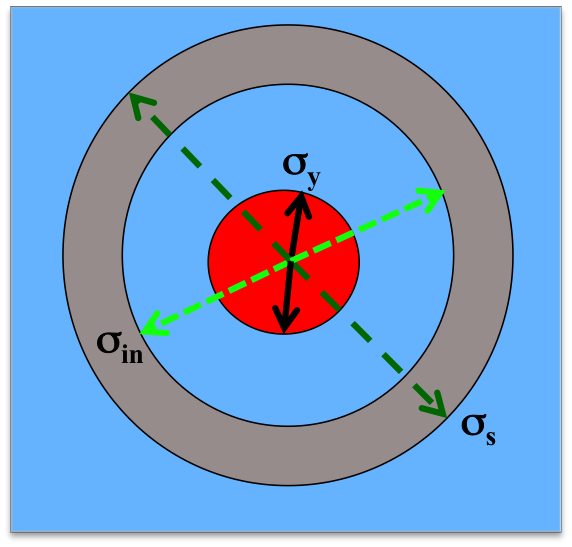}
\caption{\small    A schematic representation of a yolk-shell
particle. A mobile core of diameter $\sigma_y$ is trapped inside a
hollow spherical shell with inner diameter $\sigma_{in}$ and outer
$\sigma_s$. } \label{md}
\end{figure}

In this work we apply this algorithm to the simplest model
representation of a yolk-shell suspension, namely, a liquid of
Brownian hollow spherical particles (shells), each bearing one
smaller solid sphere (yolk), according to the schematic
representation in Figure 1. The interactions between all (shell
and yolk) particles will be represented by purely repulsive
hard-body potentials. Thus, the shell-shell potential is
\begin{equation}
u_{ss}(r)=\left\{ \begin{array}{ll}
\infty & \textrm{for $r<\sigma_{s}$} \\
0 & \textrm{for $r> \sigma_{s}$},
\end{array} \right.
\label{hss}
\end{equation}
with $\sigma_s$ being the outer diameter of the shells, the
yolk-yolk interactions are non-existent, $u_{yy}(r)=0$, and the
yolk-shell interaction is defined by
\begin{equation}
u_{ys}(r)=\left\{ \begin{array}{ll}
\infty & \textrm{for $r>\sigma_{ys}\equiv \frac{\sigma_{in} - \sigma_y}{2}$} \\
0 & \textrm{for $r< \sigma_{ys}$},
\end{array} \right.
\label{hys}
\end{equation}
where $\sigma_y$ is the diameter of the yolks and $\sigma_{in}$ is
the inner diameter of the shells.

The simulations were performed in a cubic simulation box. The
minimum image convention and periodic boundary conditions were
employed \cite{allen}. The initial configurations were generated
using the following procedure. First, particles were placed
randomly in the simulation box at the desired density and then the
overlap between the particles are reduced or eliminated. Once the
initial configuration is constructed, several thousand  cycles are
performed to lead the systems to equilibrium, followed by at least
two million cycles where the data are collected. Throughout this
paper we shall take $\sigma_s$ and $\sigma^2_s/D^0_s$ as the units
of length and time, respectively. In reality, the BD algorithm
above is only defined for systems with continuous pair potentials.
Thus, in practice we employ a softened version of the potential
above to describe the interactions among yolk and shell particles.
More specifically, we represent the hard-sphere interactions in
Eqs. (\ref{hss}) and (\ref{hys}), by inverse power-law (IPL)
potentials with amplitudes chosen such that, when written in units
of $k_BT$, read
\begin{equation}
u_{ss}(r)=(\sigma_s/r)^{12}
\end{equation}
or in the case of yolks, which only  interact with their own
shell, is given as \begin{equation}
u_{ys}(r)=[\sigma_{y}/(\sigma_{ys}-r)]^{12}
\end{equation}
where $\sigma_{ys}=(\sigma_{in}-\sigma_{y})/2$ and $\sigma_{in}$ is
the inner diameter of the shell.

In order for the properties of this soft system to provide an
accurate quantitative representation of the properties of the
original hard system, one has to provide a precise prescription
that assigns an equivalent soft system to any given hard system.
Such prescription amounts to determine the inner shell diameter
$\sigma^{eff}_{in}$ of the effective soft system by the ``blip
function'' condition, that the integral $\int \exp [-u_{ys}(r)]
d^3r$ of the effective and the original hard systems coincide, and
to determine the effective outer shell diameter $\sigma^{eff}_s$
with the analogous blip function condition between the effective
soft and the real hard shell-shell potentials. In practice,
however, we actually employed a more precise determination of
$\sigma^{eff}_s$, in which (the second maximum of) the radial
distribution function $g_{ss}(r)$ of the soft-sphere system
coincides with that of the given HS system, as discussed in detail
for solid spheres in Refs. \cite{guev, lety}. To illustrate this
procedure, figure 2 plots the simulated $g_{ss}(r)$ of the soft
system (symbols) with the $g_{ss}(r)$ corresponding to the hard
system  (solid line) at shell volume fraction $\phi=0.3$. The
latter is provided by the Percus-Yevick
 \cite{percus} approximation with its Verlet-Weis correction
\cite{verlet} (approximation denoted here as PYVW) for the liquid
of hard spheres at volume fraction $\phi=0.3$. The corresponding
PYVW static structure factor $S(k)$ will also be employed as the
structural input required by the SCGLE theory presented below.
Having established the equivalence of hard and soft sphere
systems, from now we will only refer to the shell volume fraction
$\phi$ of the hard system.

\begin{figure}
\includegraphics[width=0.5\textwidth]{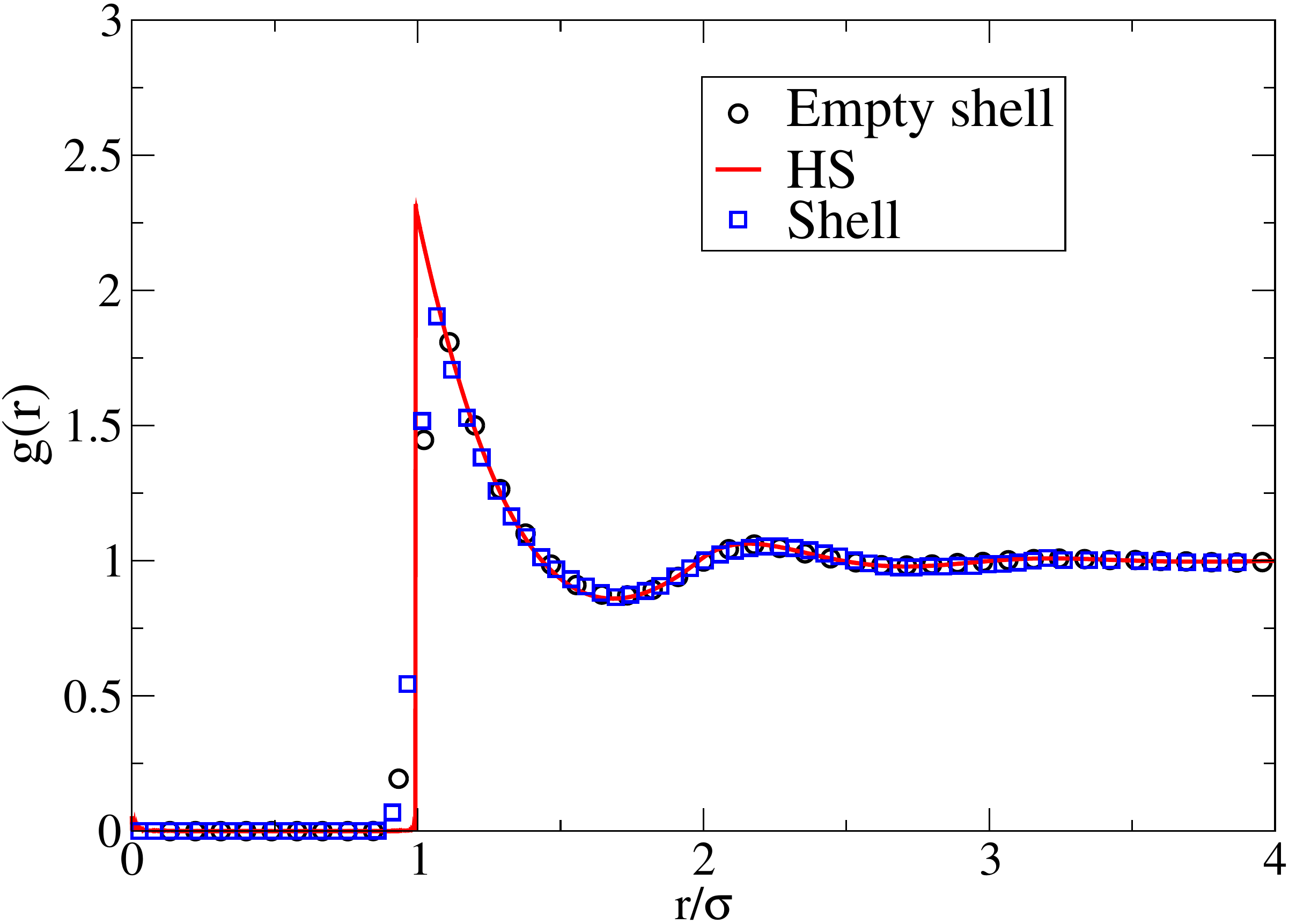}
\caption{\small Radial distribution function for shell-shell of the yolk-shell
liquid (squares) and empty shell liquid (circles).
Solid line corresponds to the Percus-Yevick approximation for HS
with the Verlet-Weis correction a volume fraction $\phi=0.3$.}
\label{hrs}
\end{figure}

The dynamic properties of the system above were then calculated
from the equilibrium configurations generated by the BD
simulations just described. Time correlation functions, like the
mean-squared displacement (MSD) $ W(t) = <[\Delta
\vec{r}(t)]^2>/6$ and the self-intermediate scattering function (self-ISF)
$F_S(k,t)$ were calculated using the efficient, low-memory
algorithm proposed in Ref. \cite{dub}. Later on in the paper these
results will be discussed and employed to assess the numerical
accuracy of the SCGLE theory for the yolk-shell systems developed
in the following section.

\section{SCGLE theory for Yolk-Shell suspensions.}

In this section we outline the derivation of a first-principles
statistical mechanical description of the dynamic properties of
our model yolk-shell suspension. We approach this task in the
framework of the self-consistent generalized Langevin equation
(SCGLE) theory of colloid dynamics\cite{scgle0, scgle1, scgle2,
rmf, todos1,todos2}, which is thus adapted here to the context of
a suspension of Brownian yolk-shell particles. The microscopic
dynamics represented by the $2N$ Langevin equations in Eqs.
(\ref{eq1}) and (\ref{eq1y}), which was the basis for the
derivation of the BD algorithm, also constitute the starting point
of the present statistical mechanical derivations. In contrast
with the strategy followed to derive the BD algorithm, in which we
took the overdamped limit in these equations right at the outset,
in the theoretical derivation below this limit is taken at a later
stage. In fact, in the absence of the yolks, only the $N$ Langevin
equations in Eq. (\ref{eq1}), with $ u_{sy}(r)=0$, remain, and
these equations were taken as the starting point in the original
derivation of the SCGLE theory for monodisperse suspensions. Thus,
what we need now is to extend this derivation to the more general
case in which we also have to deal with the presence of the yolks,
i.e., with the terms involving the yolk-shell interactions $
u_{sy}(r)$ and the additional $N$ Langevin equations in Eq.
(\ref{eq1y}). This task, however, follows essentially the same
steps as the original derivation, and hence, here we only
summarize the main arguments and the essential results, with some
details being provided in the appendix.

Just like in the Brownian dynamics simulations, the present
derivations will also be restricted to the case in which each yolk
only interacts with its own shell. Under these conditions, Eqs.
(\ref{eq1}) and (\ref{eq1y}) simplify to
\begin{equation}
M_s\frac{d{\bf v}_{i}(t)}{dt}= -\zeta^0_s{\bf v}_{i}(t)+{\bf
f}^{0} _{i}(t)-\sum_{j\neq i}\nabla_i u_{ss}(|{\bf x}_{i}(t)-{\bf
x}_{j}(t)|) - \nabla_{{\bf x}_{i}} u_{ys}(|{\bf x}_{i}(t)-{\bf
y}_{i}(t)|) \label{eq1p}
\end{equation}
and
\begin{equation}
M_y\frac{d{\bf w}_{i}(t)}{dt}= -\zeta^0_y{\bf w}_{i}(t)+{\bf
g}^0_i(t) - \nabla_{{\bf y}_{i}}
u_{ys}(|{\bf y}_{i}(t)-{\bf x}_{i}(t)|). \label{eq1py}
\end{equation}
This means that the motion of each yolk is only coupled with the
motion of its own shell through the last term of these two
equations (involving the pair potential $u_{ys}(|{\bf
x}_{i}(t)-{\bf y}_{i}(t)|)$), while the motion of each shell is
also coupled with the motion of all the other shells through the
term involving the shell-shell pair potential $u_{ss}(|{\bf
x}_{i}(t)-{\bf x}_{j}(t)|)$.

The general strategy that we shall adopt is to first average out
the degrees of freedom (velocities and positions) of the yolk
particles from the detailed description provided by Eqs.
(\ref{eq1p}) and (\ref{eq1py}) above, which involve the degrees of
freedom of both, the shell and the yolk particles. This is
equivalent to ``solving'' Eqs. (\ref{eq1py}) for the positions
${\bf y}_{i}(t)$ and velocities ${\bf w}_{i}(t)$ of the yolk
particles, and substituting the solution in Eqs. (\ref{eq1p}).
Such  procedure, detailed in the appendix,  yields the following
set of $N$ ``renormalized'' Langevin equations involving the
positions and velocities of only the shell particles
\begin{equation}
M_s{\frac{d{\bf v}_{i}(t)}{dt}}= -\zeta^0_{s}{\bf v}_{i}(t)-
\int_0^t dt'\Delta \zeta_y(t-t')  {\bf v}_{i}(t')+{\bf F}
_{i}(t)-\sum_{j\neq i}\nabla_i u_{ss}(|{\bf x}_{i}(t)-{\bf
x}_{j}(t)|) \label{eq2ppp}
\end{equation}
for $i=1,2,\ldots ,N$, where the random force ${\bf F}_{i} (t)$
has zero mean and correlation function given by $ \langle {\bf
F}_{i}(t){\bf F}_{j}(0)\rangle =k_{B}T[\zeta^0_{s}2\delta (t)+
\Delta \zeta _y(t) ]\delta _{ij}\stackrel{\leftrightarrow }{{\bf
I}}$ (with $i,j=1,2,\ldots ,N)$, with the time-dependent friction
function $\Delta \zeta_y (t) $ given by the \textit{approximate}
result in Eq. (\ref{dzdty4}), namely,
\begin{equation}
\Delta\zeta_y(t)= \frac{k_BT n_0}{3(2\pi)^3}\int d^3 k
[kg_{ys}(k)]^2 e^{-k^2D^0_y t} F_S(k,t). \label{dzdty4s}
\end{equation}
In this equation $F_S(k,t)$ is the self intermediate scattering
function of the shell particles and $g_{ys}(k)$ is the Fourier
transform of $g_{ys}(r) \equiv\exp[-\beta u_{ys}(r)]$, i.e.,
$n_0g_{ys}(r)$  is the probability distribution function that the
center of the yolk lies a distance $r$ apart from the center of
the shell, normalized such that  $n_0\equiv1/ \int \exp[-\beta
u_{ys}(r)] d^3r$.

The second stage in this strategy is to take the set of $N$
``renormalized'' Langevin equations in Eq. (\ref{eq2ppp}) as the
starting point of the derivation of three results that are central
to the formulation of the SCGLE theory. The first is a generalized
Langevin equation for a single tracer shell particle, which
effectively takes into account its interactions with the rest of the
``renormalized'' shells. This amounts to solving $(N-1)$ of the
Langevin equations in Eq. (\ref{eq2ppp}) (for, say,
 $i=2, ...,
N$), and to use the result in Eq. (\ref{eq2ppp}) for $i=1$. This
leads to the desired generalized Langevin equation which, with the
label $i=1$ changed to $``T"$ (for ``tracer''), reads
\begin{equation}
M_s{\frac{d{\bf v}_{T}(t)}{dt}} =  -\zeta ^0_{s}{\bf v}_{T}(t)-
\int_0^t dt'\Delta \zeta_y(t-t')  {\bf v}_{T}(t')- \int_0^t
dt'\Delta \zeta_s(t-t')  {\bf v}_{T}(t') +{\bf F} (t). \label{re}
\end{equation}
The random force ${\bf F} (t)$ in this equation has zero mean and
correlation function given now by $ \langle {\bf F}(t){\bf
F}(0)\rangle =k_{B}T[\zeta^0_{s}2\delta (t)+ \Delta \zeta _y(t)+
\Delta \zeta _s(t) ]\stackrel{\leftrightarrow }{{\bf I}}$, with
the new time-dependent friction function $\Delta \zeta_s (t) $
representing the friction on the tracer shell particle due to its
direct interactions with the other shells.

Except for the presence of the additive yolk friction term $-
\int_0^t dt'\Delta \zeta_y(t-t')  {\bf v}_{T}(t')$, the derivation
of this equation follows step by step the derivation originally
carried out for Brownian liquids of ``compact" (not yolk-shell)
particles \cite{faraday}. Thus, here we omit the details of such
derivation, which leads to an exact expression for $\Delta \zeta_s
(t)$. As argued in that reference, after a set of well-defined
approximations, also adopted in the present case, such an exact
expression becomes
\begin{equation} \Delta \zeta_s (t)=\frac{k_BT}{3\left( 2\pi
\right) ^{3} n}\int d {\bf k}\left[\frac{ k[S(k)-1]}{S(k)}\right]
^{2}F(k,t)F_{S}(k,t), \label{dzdt}
\end{equation}
where $S(k)$ is the (shell-shell) static structure factor, and $F
(k,t)$ and $F_S (k, t)$ are the collective and the self
intermediate scattering functions.

In order to evaluate $\Delta \zeta_y(t)$ and $\Delta \zeta_s(t)$
we thus need to determine $F (k, t)$ and $F_S (k, t)$. This task
involves the derivation of the other two general results, namely,
the (exact) memory function expressions for $F (k, t)$ and $F_S
(k, t)$, starting again from the mesoscopic description provided
by the set of $N$ ``renormalized'' Langevin equations in Eq.
(\ref{eq2ppp}). Here too, except for the presence of the additive
yolk friction terms in the right side of these equations, such
derivation follows step by step the corresponding derivation for
``solid" particles described in Ref. \cite{scgle0}, and hence, we
also omit the details. Furthermore, in the present case we adopt
the same set of approximations summarized in Ref. \cite{todos2} to
turn those exact memory function expressions into useful
approximate results (adapted here to include the presence of the
yolk friction function $\Delta \zeta_y(t)$). These are the
Vineyard-like approximation, in which one approximates the memory
function $C(k,t)$ of the collective ISF $F(k,t)$ by the memory
function $C_S(k,t)$  of the \emph{self} ISF $F_S(k,t)$, along with
the approximation that writes $C_S(k,t)$ as the superposition of
the normalized (yolk) friction function $\Delta \zeta_y^*
(t)\equiv \Delta \zeta_y (t)/\zeta^0_s$ plus the contribution due
to shell-shell interactions, factorized as the product of its long
wave-length limit $\Delta \zeta_s^* (t)\equiv \Delta \zeta_s
(t)/\zeta^0_s$ times a phenomenological ``interpolating function"
$\lambda (k)$. This results in the following approximate
expressions for $F(k,z)$ and $F_S(k,z)$,
\begin{equation}
F(k,z)=\frac{S(k)}{z+\frac{k^{2}S^{-1}(k)D^0_s }{1+\Delta
\zeta_y^* (z) +\lambda(k)\Delta \zeta_s^* (z)}}  \label{fk}
\end{equation}
and
\begin{equation}
F_S(k,z)=\frac{1}{z+\frac{k^{2}D^0_s }{1+\Delta \zeta_y^* (z)
+\lambda(k)\Delta \zeta_s^* (z)}},  \label{fks}
\end{equation}
where $\lambda (k)$ is given by  \cite{todos2}
\begin{equation}
\lambda _\alpha(k)=1/[1+( k/k_{c})]^{2}, \label{lambdadkuniform}
\end{equation}
and $k_{c}$ is an empirical cutoff wave-vector. In this paper we
shall use the value $k_{c } = 1.305 (2\pi/\sigma)$, which results
from a previous calibration \cite{overdampedatomic} of the SCGLE
theory with simulation data for the hard-sphere system, a limit
that we must recover in the absence of yolks, i.e., when $\Delta
\zeta_y^* (z)=0$.

Eqs. (\ref{dzdty4s}) and (\ref{dzdt})-(\ref{lambdadkuniform})
above constitute the extension of the SCGLE description of the
dynamics of a liquid of yolk-shell particles. The simultaneous
solution of these equations for given static structural properties
$g_{ys}(k)$ and $S(k)$ determines the time-dependent friction
functions $\Delta \zeta_y^* (t)$ and $\Delta \zeta_s^* (z)$ and
the intermediate scattering functions $F(k,t;\phi)$ and
$F^S(k,t;\phi)$. The mean squared displacement $W(t;\phi)$ then
derives from $\Delta \zeta_y^* (t)$ and $\Delta \zeta_s^*
(t;\phi)$ according to
\begin{equation}
W(t;\phi) =D^0_st-\int_0^{t}\left[ \Delta \zeta_y^* (t-t')+\Delta
\zeta^*_s(t-t';\phi)\right]W(t';\phi)dt'.  \label{wdtforfinitephi}
\end{equation}
The solution of Eqs. (\ref{dzdty4s}) and
(\ref{dzdt})-(\ref{wdtforfinitephi}) for the yolk-shell model
system described in the previous section will now be discussed.

\section{Results}

In this section we present our theoretical and simulated results
for the properties that describe the Brownian motion of a tagged
yolk-shell complex. Our intention is to describe how the mean
squared displacement and the self intermediate scattering function
of such tagged yolk-shell particles are influenced by the combined
effect of the interaction of the shell with its own yolk and with
the other shells as we vary the concentration $n\equiv N/V$ of
yolk-shell particles, actually indicated in terms of the shell
volume fraction $\phi\equiv \pi n\sigma_s^3/6$. To simplify the
discussion, we shall begin with the description of the properties
of a single yolk-shell complex, which diffuses without interacting
with other yolk-shell particles, i.e., with the $\phi\to\infty$
limit. Under these conditions, the shell-shell friction function
$\Delta \zeta^*_s(t;\phi)$ vanishes and one only has to solve
simultaneously Eqs. (\ref{dzdty4s}) and (\ref{fks}) for $\Delta
\zeta_s^* (t)$ and $F^S(k,t;\phi)$.

After discussing this infinite dilution limit, we shall study the
additional effects that derive from the shell-shell interactions
at given finite concentrations of yolk-shell complexes. One
important aspect of our discussion in both concentration regimes,
however, will refer to the comparison of these properties with
those of a system of empty shells (i.e., of the same system but in
the absence of the yolk particles). This simpler system is
actually identical to an ordinary suspension of hard-sphere
particles without hydrodynamic interactions, whose properties will
be denoted as $W_0(t;\phi)$ and $F^S_0(k,t;\phi)$. Theoretically,
these properties are obtained from the solution of Eqs.
(\ref{dzdt})-(\ref{wdtforfinitephi}) with $\Delta \zeta_y^*
(t)=0.$

Finally, in the last section, we discuss the extension of the SCGLE theory to consider other, more complex conditions. Here we  consider the case when the
 interaction between shells were to be attractive rather than repulsive, but in which the interaction
 of the yolk-shell otherwise remains the same. We then also compare the results of yolk-shell complex
 with a system of empty shells as function of temperature.

\subsection{Yolk-shell self-diffusion in the dilute limit}

\begin{figure}
\includegraphics[width=0.45\textwidth]{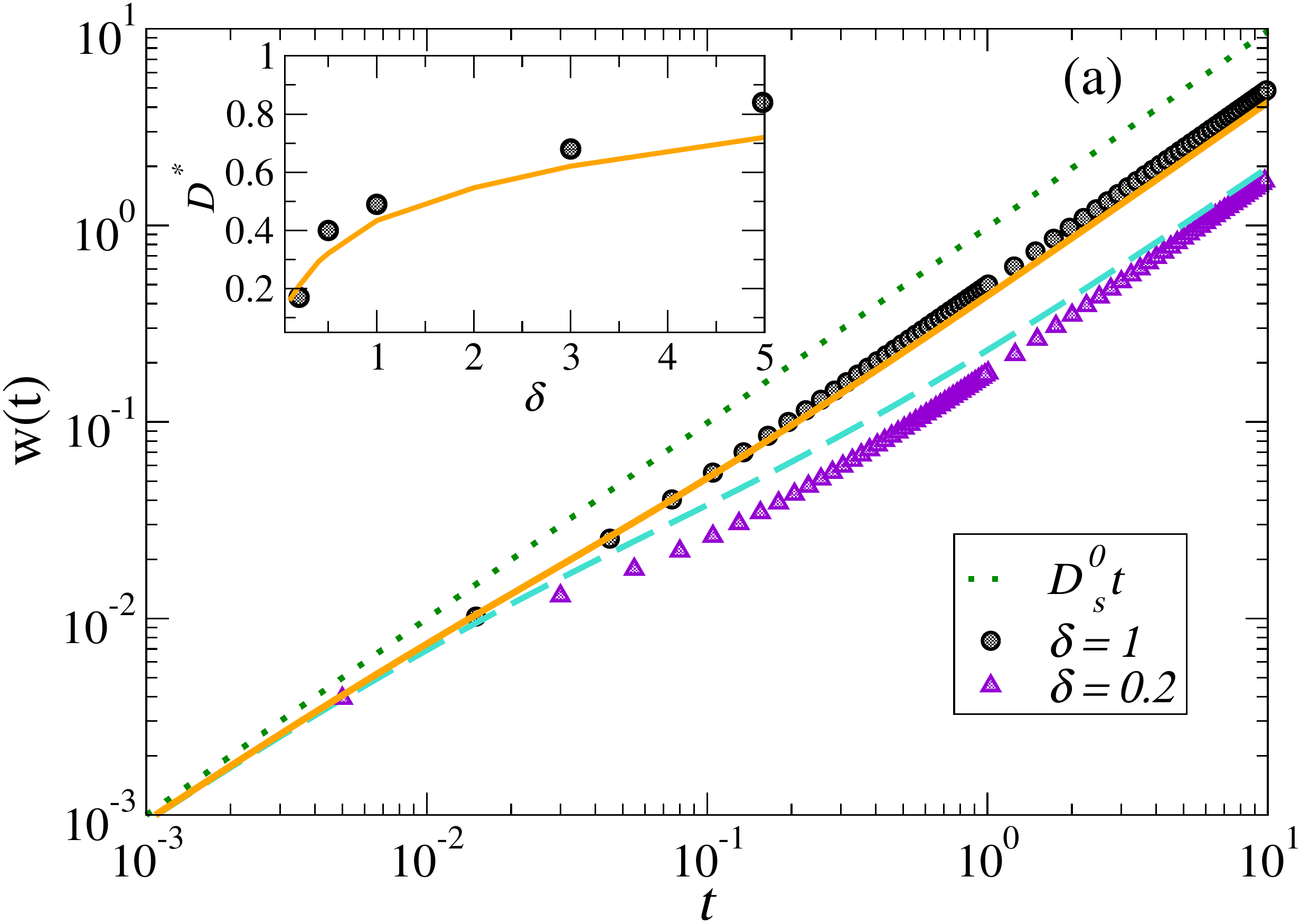}
\includegraphics[width=0.45\textwidth]{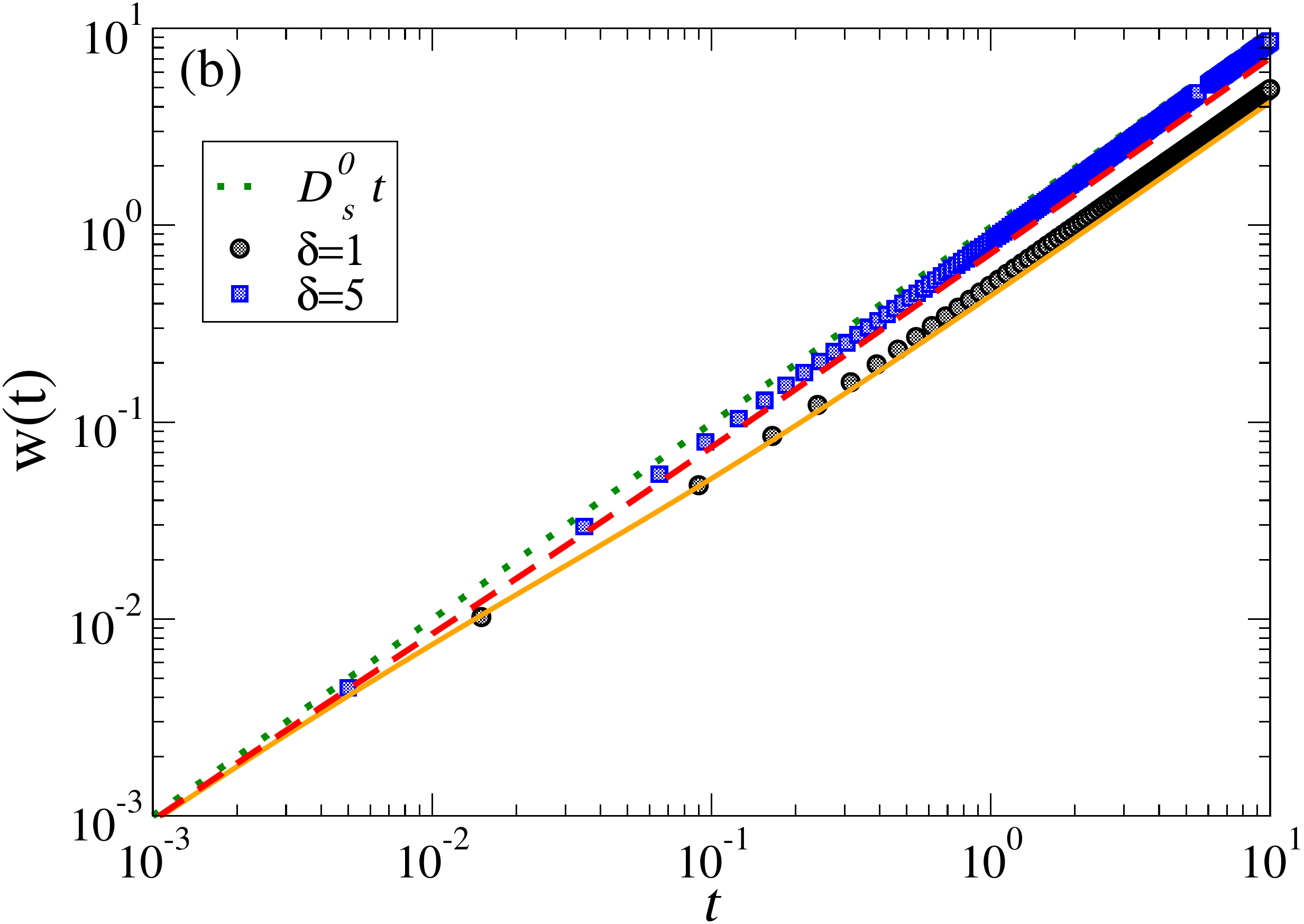}
\caption{\small  Mean square displacement $W(t;\phi=0)$ of
non-interacting yolk-shell particles with shell thickness
$(\sigma_s-\sigma_{in})/2=0.05$ (or $\sigma_{in}=0.9$) and yolk
diameter $\sigma_y=0.2$. We recall that we are taking $\sigma_s$
as the units of length and $\sigma^2_s/D^0_s$ as the time unit.
The dotted line represents the MSD $W_0(t;\phi=0)=D^0_{s}t$ of a
freely diffusing empty shell. The circles are the Brownian
dynamics data and the solid line is the theoretical prediction of
the SCGLE theory corresponding to a dynamic asymmetry parameter
$\delta\equiv D^0_y/D^0_s=1$. The squares and the dashed line in
(a) correspond to $\delta=0.2$, and the triangles and dot-dashed
line in (b) correspond to $\delta=5.$ Finally the inset in figure (a)
we compare the simulation and theoretical results for the long-time
self-diffusion $D^*$ as function $\delta$. } \label{wdt1}
\end{figure}

Let  us thus start by analyzing the results of our Brownian
Dynamics simulations for the mean squared displacement
$W(t;\phi=0)$ of a freely-diffusing yolk-shell complex, which we
present in Fig ~\ref{wdt1} (symbols). Let us mention that all the
results discussed in this figure, and in general in the present
section, correspond to a fixed yolk-shell geometry, in which the
thickness $(\sigma_s-\sigma_{in})$ of the shell is 5\% the shell's
outer diameter, $\sigma_{in}/\sigma_s=0.9$, and the yolk's
diameter is 0.2 in units of $\sigma_s$, i.e.,
$\sigma_{y}/\sigma_s=0.2$. The three different symbols in Fig
~\ref{wdt1} correspond to three different values of the ratio
$\delta\equiv D^0_y/D^0_s$, which measures the dynamic asymmetry
in the short-time diffusion of the yolk and the shell particles.
These values are $\delta=0.2,\ 1,$ and 5. Let us first focus only
on the circles, corresponding to the BD simulation results for
$\delta=1$, and on the dotted line representing the MSD
$W_0(t;\phi=0)=D^0_{s}t$ of a freely diffusing empty shell. The
deviation of the simulation data from $W_0(t;\phi=0)$ is a measure
of the additional friction effects on the displacement of the
complex due to the yolk-shell interaction. The solid line that
follows the BD data in circles are the prediction of our SCGLE
theory obtained, as indicated above, by simultaneously solving
Eqs. (\ref{dzdty4s}) and (\ref{fks}) for $\Delta \zeta_s^* (t)$
and $F^S(k,t;\phi)$, and then solving Eq. (\ref{wdtforfinitephi})
with $\Delta \zeta^*_s(t;\phi=0)=0$ to obtain $W(t;\phi=0)$. The
first conclusion that we can draw from this comparison is that the
SCGLE-predicted deviation of $W(t;\phi=0)$ from $W_0(t;\phi=0)$,
coincides very satisfactorily with the deviation observed in the
BD data.

The next conclusion to draw from the results in Fig. \ref{wdt1}(a)
is that the deviation of $W(t;\phi=0)$ from $W_0(t;\phi=0)$
increases when the dynamic contrast parameter $\delta$ decreases.
This is illustrated by the comparison of the BD simulations
corresponding to $\delta=0.2$ (squares) with the BD data
corresponding to $\delta=1.0$ (circles). This means, for example,
that if the interior of the shell becomes more viscous, so that
the ratio $\delta$ decreases, then also the overall diffusivity of
the yolk-shell particle will decrease. As evidenced by the solid
and dashed lines in Fig. \ref{wdt1}(a), this trend is also
predicted by the SCGLE theory, in good quantitative agreement with
the simulation data. In fact, our theory predicts, and the
simulations corroborate, that this trend is reversed when one
considers the opposite limit, in which $\delta$ is now larger than
1, as illustrated in Fig. \ref{wdt1}(b), where we compare the
(theoretical and simulation) results corresponding to $\delta=5$
(triangles and dashed line) with the previous results for
$\delta=1$ (circles and solid line).

As illustrated by the results in Fig. \ref{wdt1}, the MSD of the
yolk-shell complexes exhibits two linear regimes. At short times $
W(t) \approx D^0_s t$  and at long times $W(t) \approx D_Lt$,
where $D_L$ is the long-time self-diffusion coefficient. Thus,
$D_L$, normalized as $D^*(\phi)\equiv D_L/D_s$, is obtained from
the long-time slope of the msd, i.e, $D^*(\phi)=\lim_{t \to
\infty} <W(t)>/D^0_st$. The results for $D^*(\phi=0)$, obtained
from the predicted and simulated results for $W(t;\phi=0)$ in this
figure corresponding to different values of the dynamic asymmetry
parameter $\delta$, are summarized in the inset of Fig.
\ref{wdt1}(a), which plots $D^*(\phi=0)$ as a function of
$\delta$. There we can see that the reduction of the mobility
$D^*(\phi=0)$ from its unit value in the absence of yolks, may be
considerable. For example, for $\delta=0.2$ we have that $D^*
\approx 0.17$. A reduction of $D^*(\phi)$ of a similar magnitude
in a system in which  yolk-shell interactions have been suppressed
(i.e., in the absence of yolks) can also be produced as a result
of pure shell-shell interactions, but only at shell volume
fractions above 40\%. This can be concluded from the results of the inset
in Fig. \ref{fs1}b of the following section, which discusses the
effects of shell-shell interactions.

\begin{figure}
\includegraphics[width=0.45\textwidth]{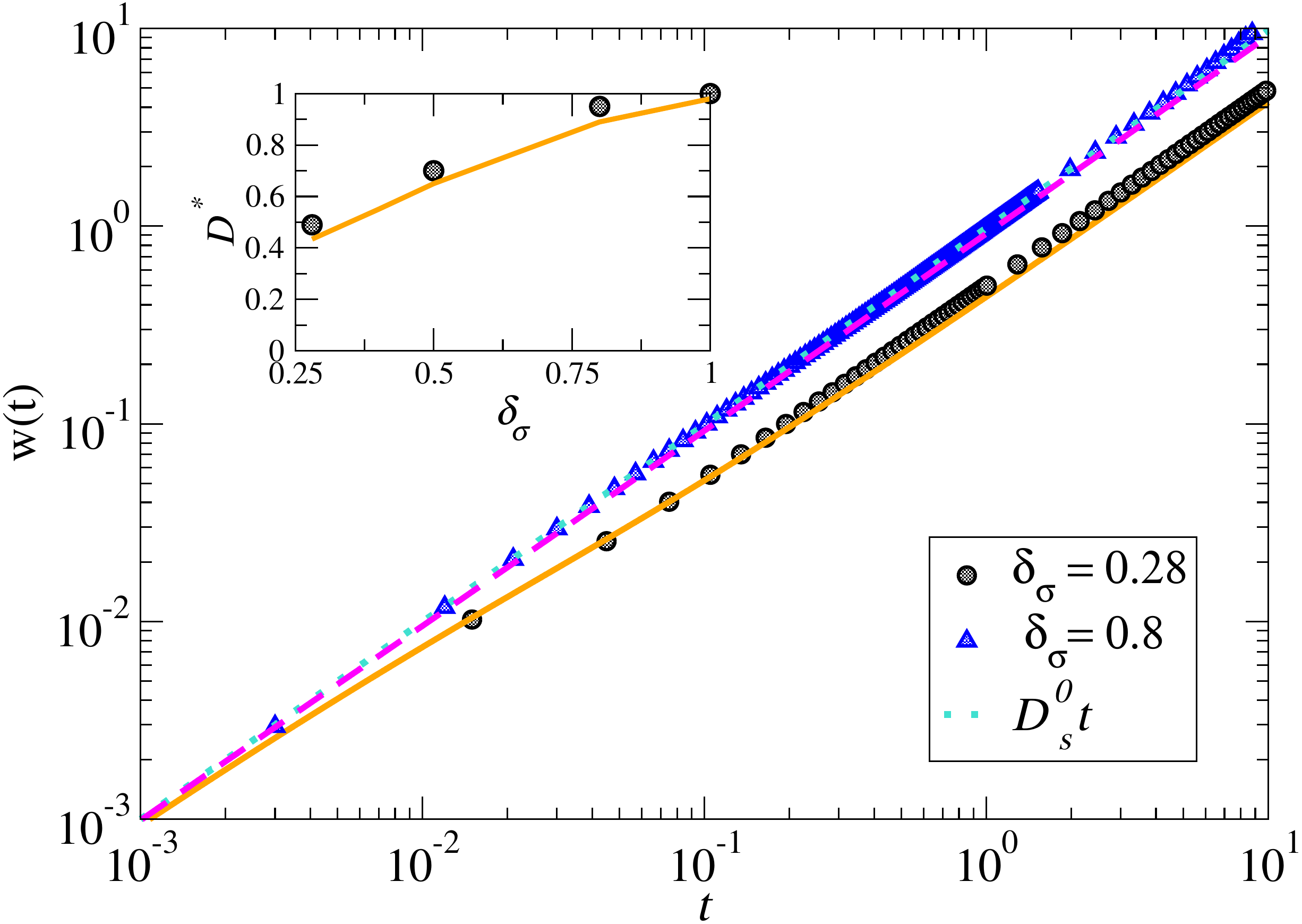}

\caption{\small  Mean square displacement $W(t;\phi=0)$ of
non-interacting yolk-shell particles with shell thickness
$(\sigma_s-\sigma_{in})/2=0.05$ (or $\sigma_{in}=0.9$) as a function of the
$\delta_\sigma \equiv \sigma_{y}/2\sigma_{ys} $. We recall that we are taking $\sigma_s$
as the units of length and $\sigma^2_s/D^0_s$ as the time unit.
The dotted line represents the MSD $W_0(t;\phi=0)=D^0_{s}t$ of a
freely diffusing empty shell. The circles are the Brownian
dynamics data and the solid line is the theoretical prediction of
the SCGLE theory corresponding to a $\delta_\sigma=0.28$.
 The triangles and the dashed line
 correspond to $\delta_\sigma=0.8$. Finally the inset in figure
we compare the simulation and theoretical results for the long-time
self-diffusion $D^*$ as function $\delta_\sigma$. } \label{wtsize}
\end{figure}

An additional important factor for the yolk-shell diffusion could be the effect of the
geometry of the particle. For example, the effect of the size of the yolk.
Fig. \ref{wtsize} shows the mean square displacement when we increase the size of the yolk,
measured using the ratio $\delta_\sigma \equiv \sigma_{y}/2\sigma_{ys} $ with fixed inner
diameter $\sigma_{in}=0.9$ and  $\delta\equiv D^0_y/D^0_s=1$.  Circles correspond to the previous
results shown in figure \ref{wdt1}, in which the size of the yolk was $\sigma_y=0.2$ ($\delta_\sigma=0.28$)
with $\delta\equiv D^0_y/D^0_s=1$.  Also the dotted line
represents the MSD $W_0(t;\phi=0)=D^0_{s}t$ of a freely diffusing empty shell.  If we
increase the size of the yolk until the inner space is filled 80$\%$, i.e. $\delta_\sigma=0.8$,
we notice the deviation between $W(t;\phi=0)$ and $W_0(t;\phi=0)$ begins to more closely
resemble the diffusion of  the empty shell.
This is illustrated by the comparison of the BD simulations
corresponding to $\delta_\sigma=0.8$ (triangles) with the BD data
corresponding to $\delta_\sigma=0.28$ (circles). This suggests that,
when the size of the yolk increases, the yolk has less space for diffusion, and the
additional friction effects upon the displacement of the
complex due to the yolk-shell interaction decrease. The effect is similar to that
found when the inside the yolk diffuses in medium
less viscous than the shell. As evidenced by the solid
and dotted lines in Fig. \ref{wtsize}, this trend is also
predicted by the SCGLE theory, showing good quantitative agreement with
the simulation data. Finally the inset of Fig.
\ref{wtsize} plots $D^*(\phi=0)$ as a function of
$\delta_\sigma$. There we can see the reduction of the mobility
$D^*(\phi=0)$ when we decrease the size of the yolk. For
 example, for $\delta_\sigma=0.5$, we have that $D^*
\approx 0.65$.

\subsection{Yolk-shell diffusion at finite concentration}

\begin{figure}
\includegraphics[width=0.45\textwidth]{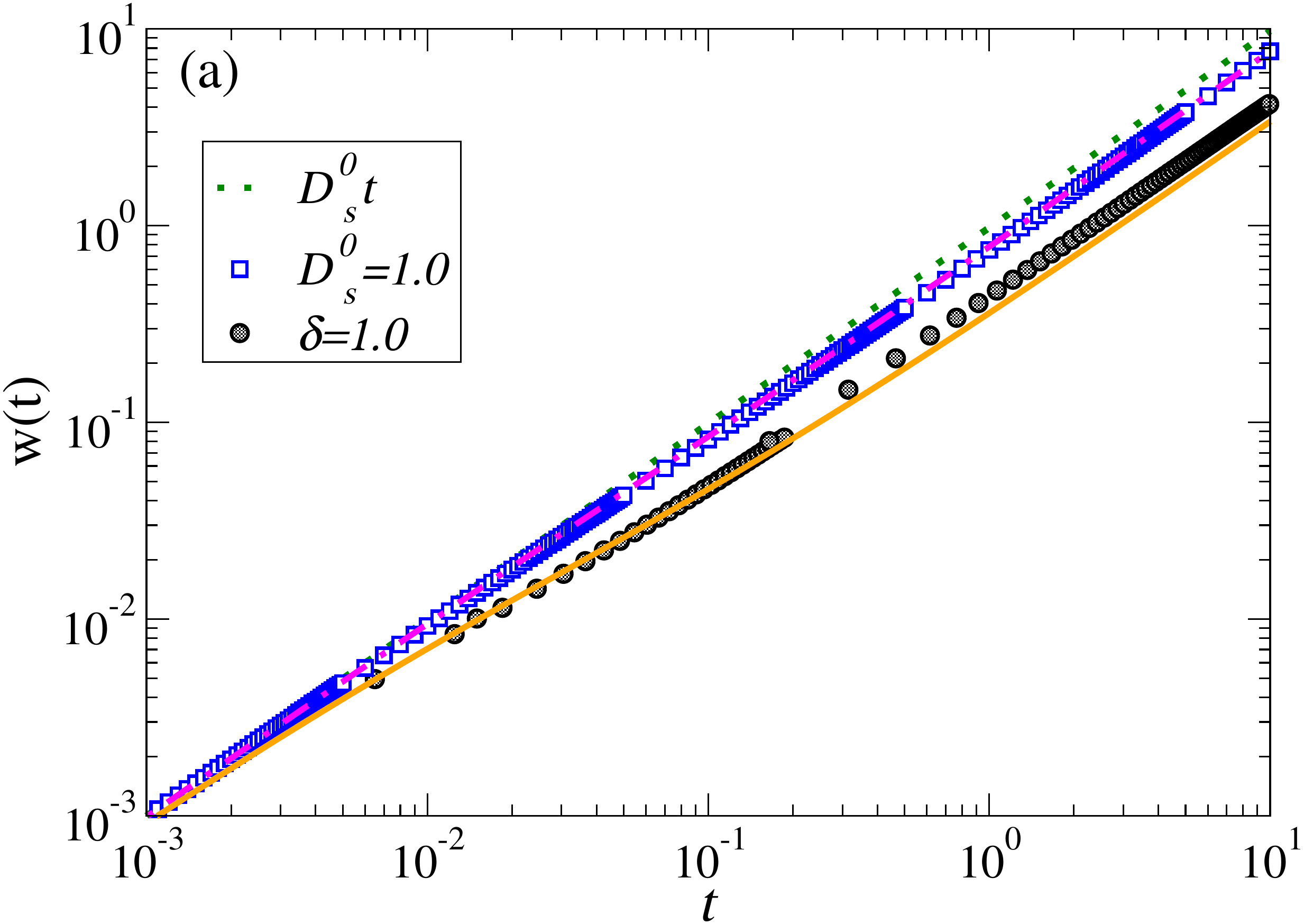}
\includegraphics[width=0.45\textwidth]{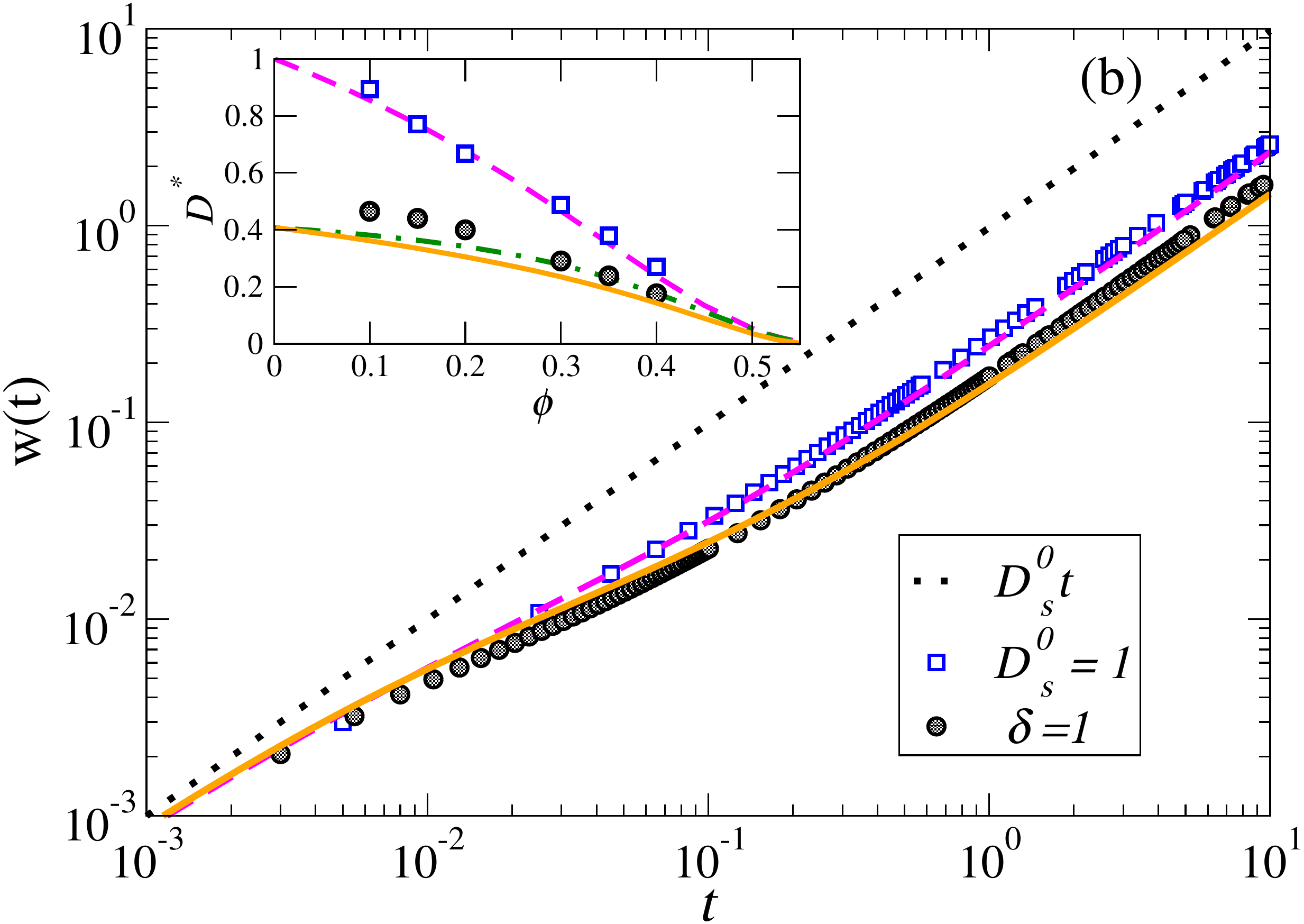}
\caption{\small Mean square displacement $W(t;\phi=0.15)$ (a) and $W(t;\phi=0.4)$ (b)  of
interacting yolk-shell particles with shell thickness
$(\sigma_s-\sigma_{in})/2=0.05$ (or $\sigma_{in}=0.9$) and yolk
diameter $\sigma_y=0.2$. We recall that we are taking $\sigma_s$
as the units of length and $\sigma^2_s/D^0_s$ as the time unit.
The dotted line represents the MSD $W_0(t;\phi=0)=D^0_{s}t$ of a
freely diffusing empty-shell. The circles are the Brownian
dynamics data for the yolk-shell  and the solid line is the theoretical prediction of
the SCGLE theory corresponding to a dynamic asymmetry parameter
$\delta=1$. The squares are the Brownian
dynamics data of the empty-shell and the dashed line is the theoretical prediction of
the SCGLE theory with $D^0_s=1$.  Finally the inset in figure (b)
we compare the simulation data and theoretical results
 for the long-time self-diffusion $D^*$ as function $\phi$. } \label{fs1}
\end{figure}

Let us now study the effects of shell-shell interactions on the
Brownian motion of tagged yolk-shell particles, which for
simplicity were suppressed in the previous discussion. These
effects manifest themselves only at finite concentrations. Thus,
let us begin by analyzing the results of our simulations in Fig.
\ref{fs1}a, where the circles represent the BD results for the MSD
$W(t;\phi=0.15)$ of  yolk-shell particles with dynamic asymmetry
$\delta=1$ at a finite but low volume fraction $\phi=0.15$. These
results are to be compared with the squares, which correspond to
the BD results for the MSD $W_0(t;\phi=0.15)$ of a liquid of empty
shells (or solid hard-spheres) at the same volume fraction
$\phi=0.15$, and which diffuse with the same short-time
self-diffusion coefficient $D^0_s$ as the shells in the yolk-shell
system. The MSD $W_0(t;\phi=0)=D^0_{s}t$ of a freely diffusing
empty shell is also plotted for reference as a dotted line.
Clearly, at this low volume fraction the MSD $W_0(t;\phi=0.15)$ of
the hard-sphere liquid is very similar to
$W_0(t;\phi=0)=D^0_{s}t$. In contrast, the yolk-shell interactions
present in the yolk-shell system leads to a MSD $W(t;\phi=0.15)$
that clearly deviates from the empty shell result.

Upon increasing the volume fraction to $\phi=0.4$ (Fig.
\ref{fs1}b), we observe that the deviation of $W(t;\phi=0.4)$ from
$W_0(t;\phi=0.4)$ remains rather similar to the corresponding
deviation observed at $\phi=0.15$ in Fig. \ref{fs1}a. We also
observe that both,  $W(t;\phi=0.4)$ and $W_0(t;\phi=0.4)$, now
deviate dramatically from the free-diffusion limit
$W_0(t;\phi=0)=D^0_{s}t$. This means that for this concentration
the mutual friction effects due to shell-shell interactions
overwhelm the ``internal'' friction effects due to yolk-shell
interactions. From the long-time BD data for $W(t;\phi=0.4)$ and
$W_0(t;\phi=0.4)$ in Fig. \ref{fs1}b we can extract the value of
$D^*(\phi)$ and $D^*_{HS}(\phi)$, which represent the mobility of
a tracer yolk-shell particle and of an empty shell, respectively.
The results are plotted in the inset of this figure, together with
the corresponding BD results for $D^*(\phi)$ and $D^*_{HS}(\phi)$
at other volume fractions (circles and squares, respectively).
This inset thus summarizes the main trends illustrated by the
results in Fig. \ref{fs1}, by evidencing  that at low volume
fractions the difference between the mobility of a yolk-shell
complex and the mobility of an empty shell, is determined only by
the yolk-shell friction, whereas at higher concentrations it is
dominated by the shell-shell interactions.

Let us notice now that the solid lines in Fig. \ref{fs1} represent
again the predictions of the SCGLE theory for the properties of
the yolk-shell system, whereas the dashed lines are the
corresponding predictions for the empty-shell (or hard-sphere)
suspension. Once again, the agreement with the simulation results
is also quite reasonable for a theory with no adjustable
parameters. Beyond this quantitative observation, however, the
theoretical description provides additional insights on the
interpretation of the qualitative trends exhibited by the
simulation data of the long-time self-diffusion coefficients
$D^*(\phi)$ and $D^*_{HS}(\phi)$. To see this, let us define
$\Delta \zeta^*_y\equiv \int^{\infty}_0 dt\Delta \zeta^*_y(t)$ and
$\Delta \zeta^*_s(\phi) \equiv \int^{\infty}_0 dt\Delta
\zeta^*_s(t;\phi)$, and let us write Einstein's relation for $D_L$
as
\begin{equation}\label{dlphi}
D^*(\phi) = \frac{1}{1 + \Delta\zeta_y^* + \Delta\zeta_s^*(\phi)},
\end{equation}
or, in terms of the low-density limit $D^*_0\equiv D^*(\phi=0) =
[1 + \Delta\zeta_y^*]^{-1}$, as
\begin{equation}\label{dl0}
D^*(\phi) = \frac{D^*_0}{1+D^*_0 \Delta\zeta^*_s}.
\end{equation}
Since for an empty shell (or hard-sphere) $\Delta\zeta_y^*=0$, so
that $D^*_0=1$, we have that the empty-shell version of this
equation is
\begin{equation}\label{dlhsphi}
D^*_{HS}(\phi) = \frac{1}{1 + \Delta\zeta_s^{*HS}(\phi)}.
\end{equation}
Thus, if we now assume that the shell-shell friction on a
yolk-shell particle is comparable to the friction
$\Delta\zeta^{*HS}_s$ in an empty-shell suspension, we may combine
Eqs. (\ref{dl0}) and (\ref{dlhsphi}) to write an approximate
relationship between $D^*(\phi)$ and $D^*_{HS}(\phi)$, namely,
\begin{equation}\label{dleyhs}
D^*(\phi) = \frac{D^*_0\times D^*_{HS}(\phi)}{D^*_{HS}(\phi)+
D^*_0[1-D^*_{HS}(\phi)]}.
\end{equation}
This expression interpolates $D^*(\phi)$ between its exact low and
high concentration limits $D^*_0$ and $D^*_{HS}(\phi)$, and the
dot-dashed line in the inset of Fig. \ref{fs1} is the result of
using this approximate expression.

\begin{figure}
\includegraphics[width=0.45\textwidth]{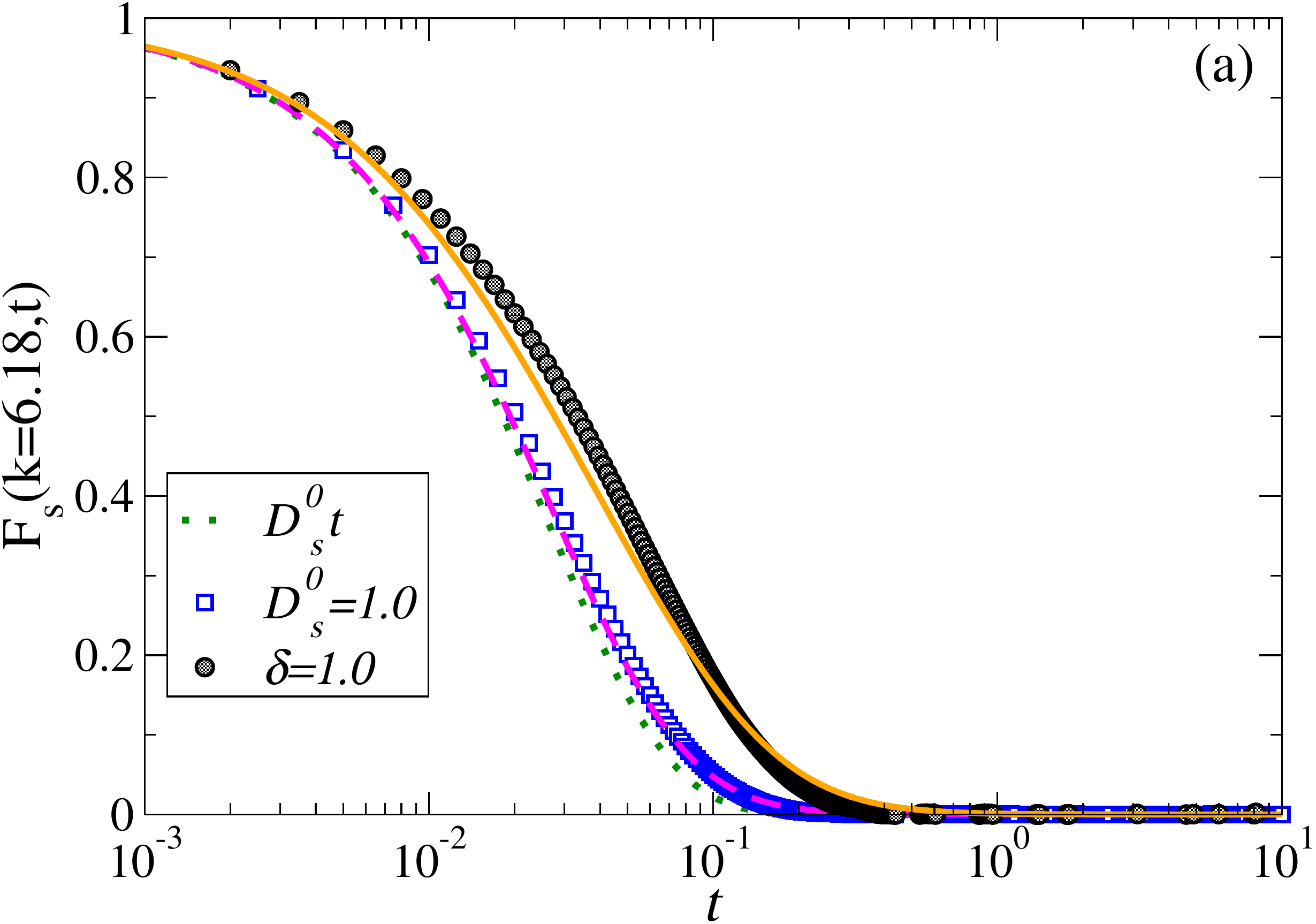}
\includegraphics[width=0.45\textwidth]{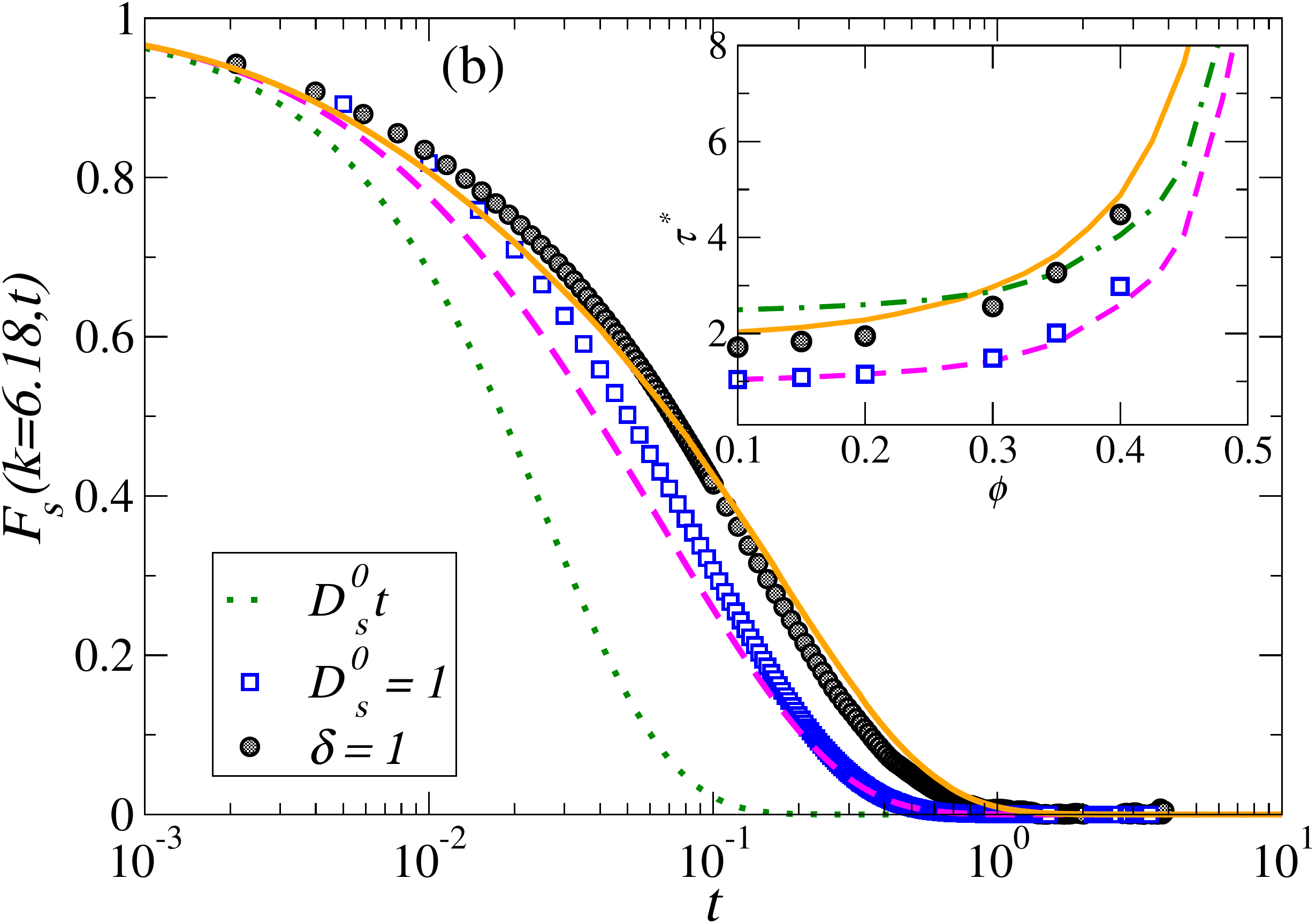}
\caption{\small The self-intermediate  $F^s(k,t;\phi=0.15)$ (a) and $F^s(k,t;\phi=0.4)$  (b)   of
interacting yolk-shell particles with shell thickness
$(\sigma_s-\sigma_{in})/2=0.05$ (or $\sigma_{in}=0.9$) and yolk
diameter $\sigma_y=0.2$. We recall that we are taking $\sigma_s$
as the units of length and $\sigma^2_s/D^0_s$ as the time unit.
The dotted line represents the self-ISF $F^s_0(k,t;\phi=0)=exp(-k^2 D^0_{s}t)$ of a
freely diffusing empty shell. The circles are the Brownian
dynamics data for the yolk-shell  and the solid line is the theoretical prediction of
the SCGLE theory with $\delta=1$. The squares are the Brownian
dynamics data of the empty-shell and the dashed line is the theoretical prediction of
the SCGLE theory  with $D^0_s=1$.  Finally the inset in figure (b) we compare the
simulation data and theoretical results for the $\alpha$-relaxation time $\tau^*\equiv
k^2D_s^0\tau_\alpha$ as function $\phi$.  } \label{fs2}
\end{figure}

We next analyze the results for the self-intermediate scattering function of shell-shell
$F^s(k,t)$ obtained of our simulation. Let us first discuss the results  in figure
\ref{fs2}a, where the circles represent the BD results for the self-ISF
$F^s(k,t;\phi=0.15)$ of  yolk-shell particles with dynamic asymmetry
$\delta=1$ at a volume fraction $\phi=0.15$.  We compare these results with the self-ISF
$F^s_0(k,t;\phi=0.15)$ of a liquid of empty-shells at the same volume fraction and with
 short-time self-diffusion coefficient $D^0_s=1$, plotted as squares. The self-ISF $F^s_0(k,t;\phi=0)= exp(- k^2
D^0_s t)$  of a freely diffusing empty shell is also plotted for reference as a dotted line.
We evaluate the self-ISF for three cases  at $k=6.18$, which is approximately the first peak of the $S(k)$.
The relaxation of the empty shell $F^s_0(k,t;\phi=0.15)$ shows a single,
exponential-like decay similar to $F^s_0(k,t;\phi=0)$. We observe that
the relaxation of $F^s(k,t;\phi=0.15)$ suffers a deviation from
$F^s_0(k,t;\phi=0.15)$, and this is also evidence of the
additional friction effect upon the yolk-shell interaction.

Upon increasing the volume fraction to $\phi=0.4$  (Fig.
\ref{fs2}b), we observe that the deviation of $F^s(k,t;\phi=0.4)$ from
$F^s_0(k,t;\phi=0.4)$ remains rather similar to the corresponding
deviation observed at $\phi=0.15$ in Fig. \ref{fs1}a. We also
observe that both, $F^s(k,t;\phi=0.4)$ and $F^s_0(k,t;\phi=0.4)$ now
deviate dramatically from the free-diffusion limit
$F^s_0(k,t;\phi=0)$. This further corroborates with the observations of the MSD. The
theoretical predictions for the liquid of yolk shell (solid line)
follow with good agreement the simulation data for $\phi=0.15$ and
show slight deviation for $\phi=0.4$. The empty-shell (dashed line)
also shows good agreement.

 The difference between the yolk-shell and empty-shell
results can also be expressed more economically in terms of the
corresponding $\alpha$-relaxation times $\tau_\alpha$, defined by
the condition $F_S(k,\tau_\alpha)=1/e$ and scaled as $\tau^*\equiv
k^2D_s^0\tau_\alpha$. The inset of Fig. \ref{fs2}b exhibits the
theoretical (solid line) and simulated (circles) results for the
yolk-shell $\tau^*(k;\phi,\delta)$ evaluated at $k=6.18$ for
$\delta=1$ as a function of $\phi$. These results may be compared
with the theoretical (dashed line) and simulated (squares) results
for $\tau^*_{HS}(k=6.18;\phi=0.4)$, corresponding to the
empty-shell suspension, with similar conclusions as in Fig. \ref{fs1}b.
In analogy with the relationship in Eq. (\ref{dleyhs}), from the
SCGLE equations one can also derive an approximate relationship
between $\tau^*(k;\phi,\delta)$ and $\tau^*_{HS}(k;\phi)$, namely,
$\tau^*(k;\phi,\delta)\approx \tau^*_{HS}(k;\phi) +
\Delta\zeta_y^*(\delta)$. This prediction of the value of
$\tau^*(k;\phi,\delta)$ has a rather modest quantitative accuracy,
as indicated by the dot-dashed line in the inset. Still, it
contributes to a simple and correct qualitative understanding of
the main features of the properties of the yolk-shell system being
studied.

\begin{figure}
\includegraphics[width=0.45\textwidth]{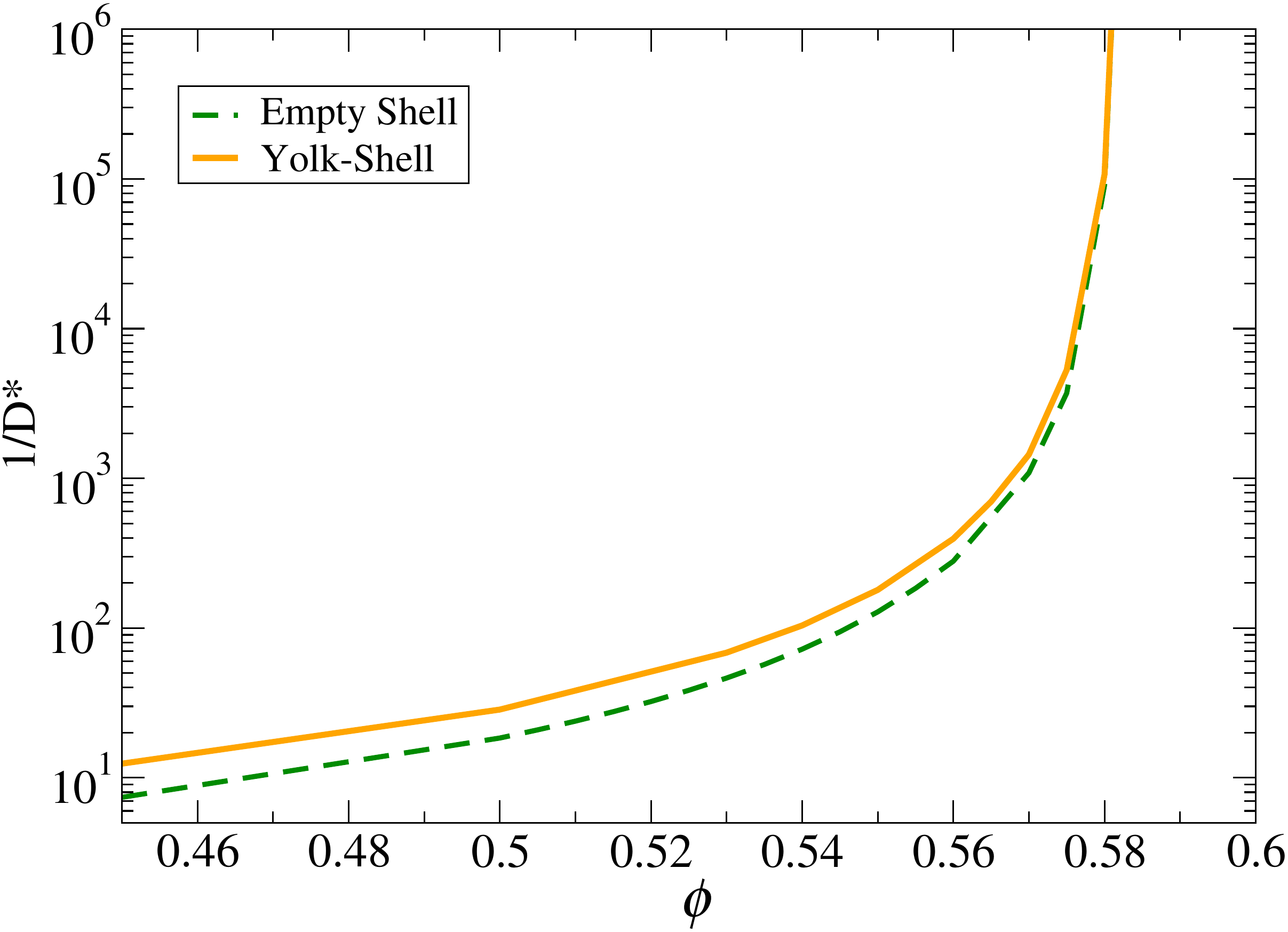}
\includegraphics[width=0.45\textwidth]{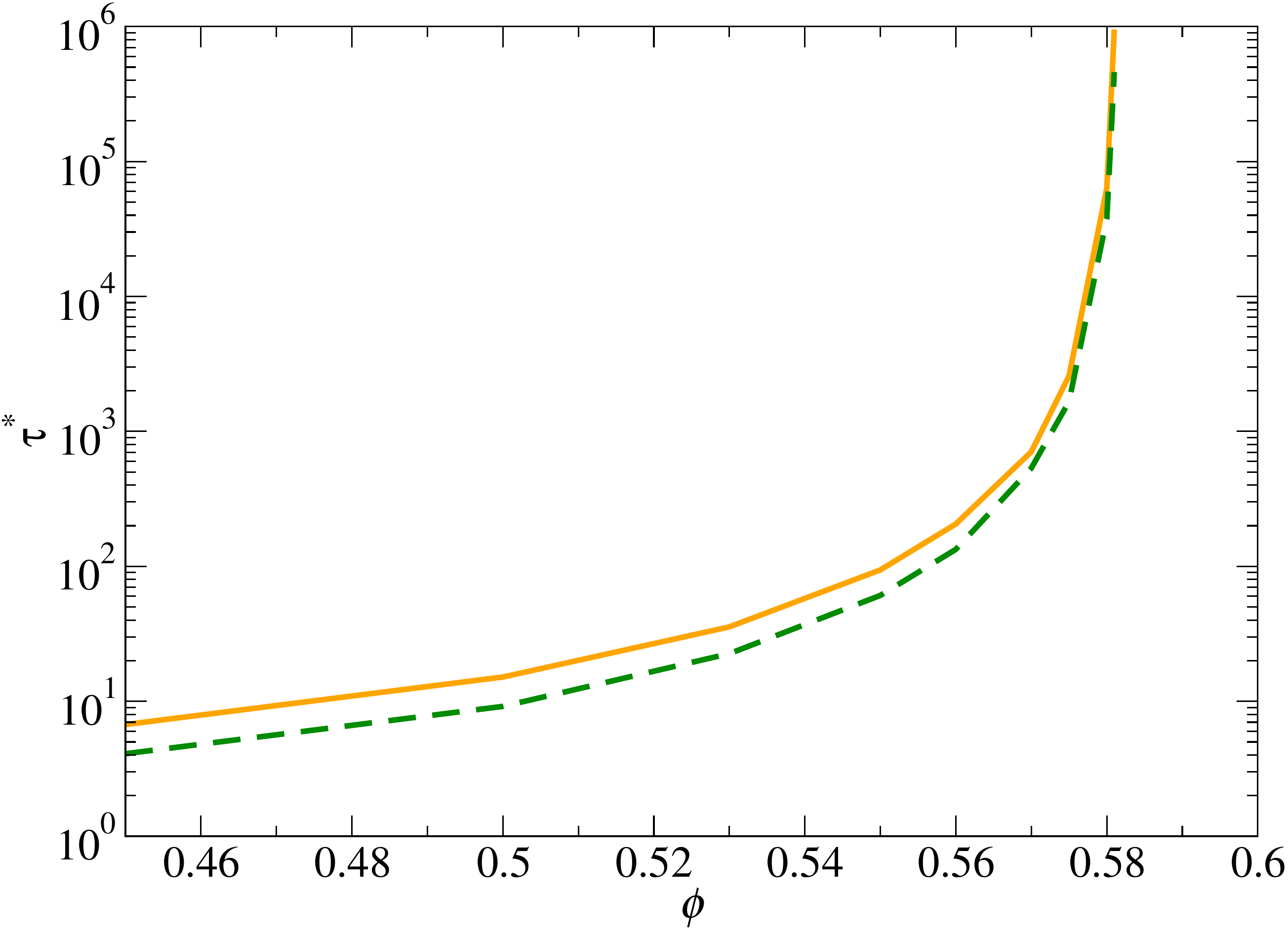}
\caption{\small The long time self-diffusion coefficient $D^*$ and
$\alpha$-$relaxation$ time $\tau^*\equiv k^2D_s^0\tau_\alpha$as
function of the volume fraction $\phi$.  The solid line is
theoretical results for the yolk-shell complex. The dashed line is
theoretical results of the empty shell.} \label{bey}
\end{figure}

The previous results demonstrated the SCGLE theory to predict,
with good agreement, the different results obtained with BD
simulations of the yolk-shell complex.  Based upon this agreement,
our theoretical approach may be extended to consider other, more
complex conditions.   For example, the previous results only
showed BD simulation data  to below a volume fraction of
$\phi=0.5$.  But here, we also demonstrate the use of the theory
to calculate dynamic properties of the yolk-shell at high volume
fractions.  This is shown in figure \ref{bey}, which plots the
inverse of the long time self-diffusion coefficient  $1/D^*$ and
$\alpha$-$relaxation$ time  $\tau^*\equiv k^2D_s^0\tau_\alpha$ as
a function of the volume fraction $\phi$ beyond $\phi=0.5$.  The
solid curve shows the theoretical prediction for the yolk-shell
complex with  $\delta=1$. The dashed line represents the
theoretical prediction for the empty shell with $D^0_s=1$. For
both figures we note that, upon increasing the volume fraction,
the difference between the yolk-shell and empty shell diffusion
progressively decreases. Beyond $\phi=0.56$,  both systems appear
to demonstrate the same dynamic behavior.  This suggests that the
shell-shell interaction is more dominant than the yolk-shell
interaction.  We observe that at approximately $\phi=0.58$, both
systems appear to show a clear divergence, indicating the dynamic
arrest transition as predicted for a hard-sphere system
(previously predicted SCGLE theory in the ref\cite{todos2}) . We
can conclude that the additional friction effect upon the
displacement of the complex due to the yolk-shell interaction is
not important when approaching the glass transition of hard
spheres.

\subsection{Effects of shell-shell attractions}

Another possible application of the SCGLE theory is to consider
the possibility that the interaction between shells include an
attractive component, besides the hard-core repulsive interaction
just described. Such interaction between shells could be described
adding a simple 'attractive Yukawa' term to the hard-sphere
potential $u_{ss}(r)$ in Eq. (\ref{hss}), which now would be given
by
\begin{equation}
\beta u_{ss}(r)=\left\{ \begin{array}{ll}
\infty & \textrm{for $r<\sigma_{s}$} \\
-K\frac{exp[-z(r/\sigma_s-1))]}{r/\sigma_s} & \textrm{for $r> \sigma_{s}$},
\end{array} \right.
\label{hssy}
\end{equation}
An ordinary suspension with attractive interactions exhibits
interesting phase behavior and structural and dynamic properties
\cite{pedro, fuchs}. Our yolk-shell system, in which the
yolk-shell interactions remains the same as that of the equation
(\ref{hys}), also exhibits similar behavior. The state space of
this system is spanned by the hard-sphere volume fraction of the
shell and by dimensionless parameters $z$ and $K$, representing
the inverse decay length and the depth of the attractive Yukawa
well. For the calculation of the static structure factor for the
shell, we can use the mean spherical approximation (MSA) described
in reference \cite{hoy}.

\begin{figure}
\includegraphics[width=0.45\textwidth]{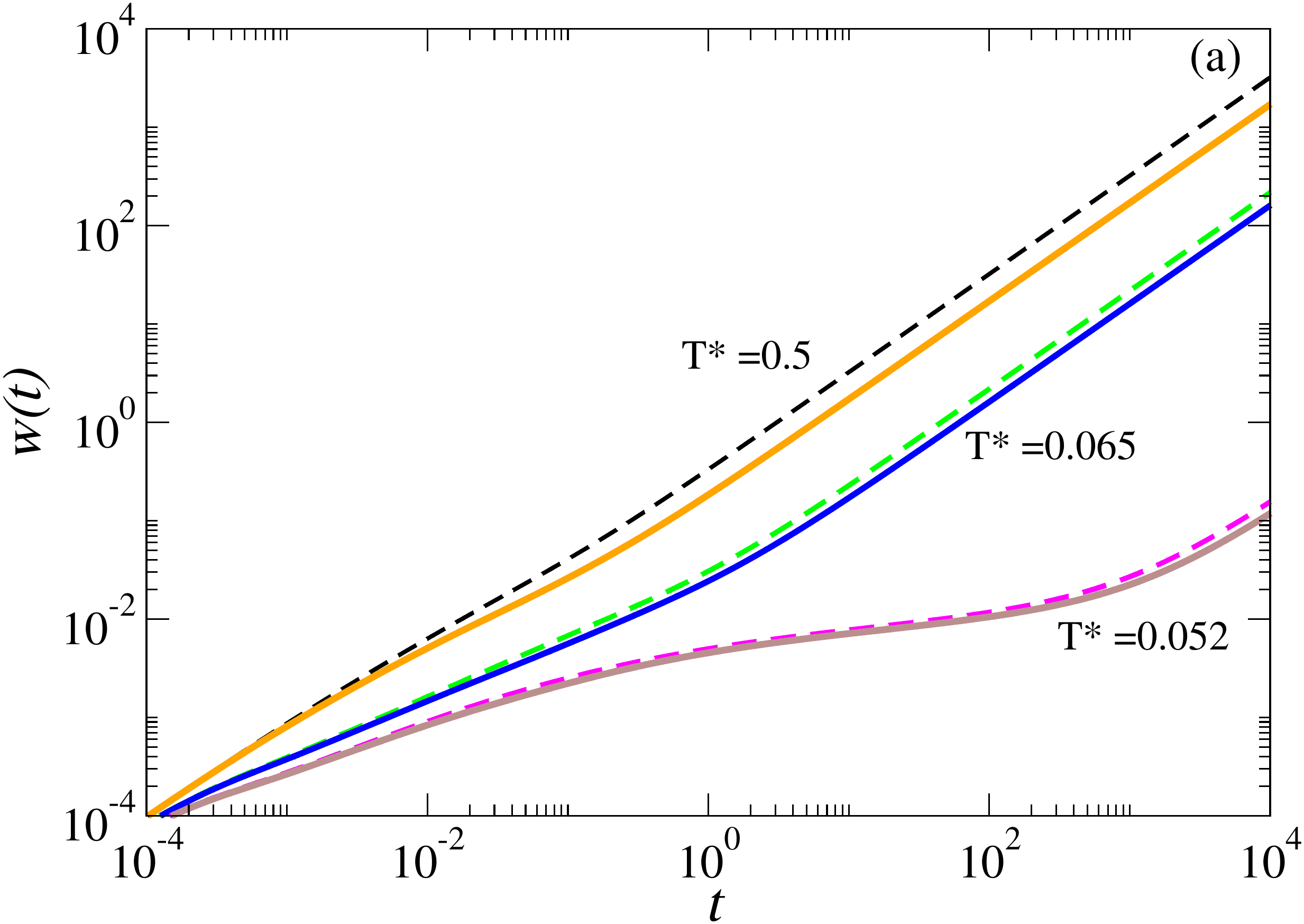}
\includegraphics[width=0.45\textwidth]{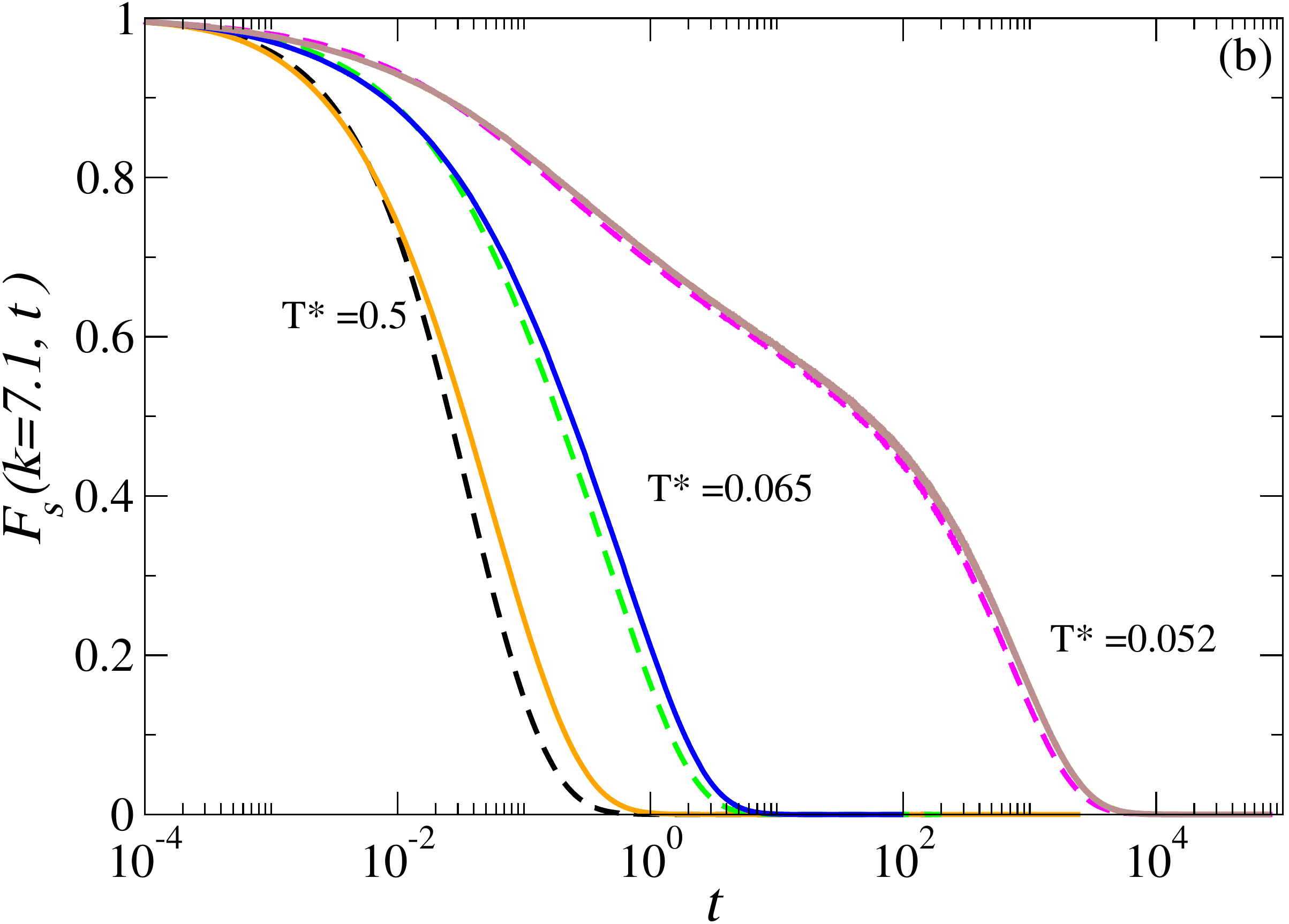}
\caption{\small The mean square displacement  and the self-intermediate  $F^s(k,t;\phi=0.35)$ as function of
 temperature $T^*=(1/K)$ with fixed $z=20$ and $\phi=0.35$. The shell thickness
$(\sigma_s-\sigma_{in})/2=0.05$ (or $\sigma_{in}=0.9$) and yolk
diameter $\sigma_y=0.2$. The solid lines is the theoretical prediction of
the SCGLE theory with $\delta=1$. The dashed lines is the theoretical prediction of
the SCGLE theory  for the empty shell with $D^0_s=1$. } \label{atrac}
\end{figure}

Let us now study the effects of shell-shell interactions on the
Brownian motion of tagged yolk-shell particles. For this case, we want to see if
the effect of attraction is influencing the dynamics of the yolk-shell complex,
and compare this with the dynamics of the empty shell, also with attractive interaction. We fixed the system with $z=20$ and $\phi=0.35$,
and lowered temperatures\ for three different temperatures $T^*=0.5$, $0.065$ and  $  0.052$.
In figure \ref{atrac}, we show the prediction of the SCGLE theory of the mean square displacement and self-intermediate scattering function
 as a function of temperature $T^*=(1/K)$.   Let us begin by analyzing the results  in Fig.
\ref{atrac}a, where the solid lines represent the prediction of SCGLE theory results for the MSD
$W(t;T^*=0.5)$ of  yolk-shell particles with dynamic asymmetry
$\delta=1$ at a temperature $T^*=0.5$. These
results are to be compared with the dash line, which correspond to
the SCGLE theory results for the MSD $W_0(t;T^*=0.5)$ of a liquid of empty
shells (or solid attractive hard-spheres) at the same temperature
$T^*=0.5$, and which diffuse with the same short-time
self-diffusion coefficient $D^0_s$ as the shells in the yolk-shell
system.  The yolk-shell  repulsive interactions is more important in this temperature than the attraction between shells.
The yolk-shell system lead to a MSD $W(t;T^*=0.5)$
that clearly deviates from the empty shell result. Upon decreasing the temperature to $T^*=0.065$ (Fig.
\ref{fs1}b), we observe that the deviation of $W(t;T^*=0.065)$ from
$W_0(t;T^*=0.065)$ is smaller  than the deviation observed at $T^*=0.5$.This means that for this temperature
the mutual friction effects due to shell-shell interactions
overwhelm the ``internal'' friction effects due to yolk-shell
interactions.   If continue decrease the temperature until   $T^*=0.052 $, we observe that the deviation of $W(t;T^*=0.052)$ from
$W_0(t;T^*=0.052)$ is not clear than  the previous temperatures.  This suggests that the shell-shell attraction is more dominant than the yolk-shell repulsive intereaction.

We next analyze the results for the self-intermediate scattering function of shell-shell
$F^s(k,t)$ obtained of our simulation. Let us first discuss the results  in figure
\ref{atrac}b, where the solid lines are the self-ISF
$F^s(k,t;T^*=0.5)$ of  yolk-shell particles with dynamic asymmetry
$\delta=1$ at a volume fraction $\phi=0.35$.  We compare these results with the self-ISF
$F^s_0(k,t;T^*=0.5)$ of a liquid of empty-shells at the same volume fraction and with
 short-time self-diffusion coefficient $D^0_s=1$, plotted as dashed line.
We evaluate the self-ISF for three cases  at $k=7.1$, which is approximately the first peak of the $S(k)$.
 We observe that
the relaxation of $F^s(k,t;T^*=0.5)$ suffers a deviation from
$F^s_0(k,t;T^*=0.5)$, and this is also evidence of the
additional friction effect upon the yolk-shell interaction. Upon decrease the temperature to $T^*=0.065$  and $T^*=0.052$  (Fig.
\ref{atrac}b). We corroborates the observations of the MSD.

\subsection{Effects of yolk-yolk repulsions}

With the SCGLE theory above one could still model many other
possible physical conditions. The previous illustrative exercise,
for example, corresponds to a realistic experiment in which one
adds a depletant agent to induce attractions between shells in a
yolk-shell suspension with hard-body interactions. One can also
imagine, instead, the possibility that the yolks are highly
charged colloidal particles encapsulated inside a rigid but
electrostatically inert shell, which only hinders its motion. If
the screening length is sufficiently large compared with the outer
diameter of the shell and with the mean interparticle distance
$d=n^{-1/3}$, one can now neglect the shell-shell interactions,
since the strong yolk-yolk repulsion will prevent direct contact
between shells. Under these conditions, we can notice that the
same theory above, with the roles of the yolks and shells
interchanged, provides a description of the dynamics of the yolks,
renormalized by the averaged effects of the motion of the shells.

\begin{figure}
\includegraphics[width=0.45\textwidth]{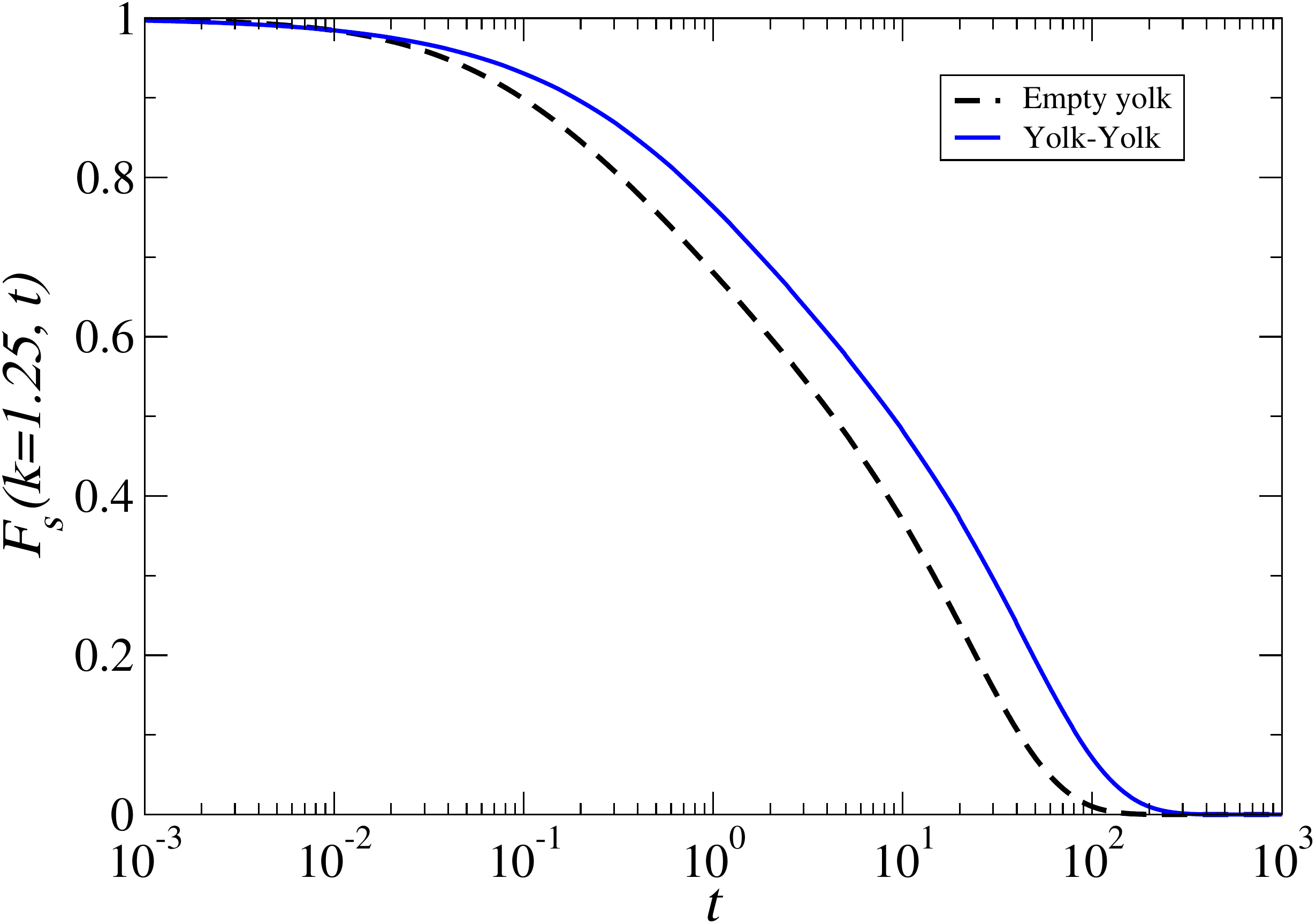}
\caption{\small Self-intermediate scattering function
$F^s(k=1.25,t)$ of the yolks in a system with  $ u_{ss}(r)=0$, $
u_{ys}(r)$ given by Eq. (\ref{hys}), and with the strong yolk-yolk
repulsions of Eq. (\ref{hyyy}) with $K=554$ and $z=0.149$. The
dashed curve corresponds to a free yolk (no shells) with the same
repulsive interactions. } \label{yolky}
\end{figure}

In Fig. \ref{yolky} we illustrate the SCGLE predictions for the
self intermediate scattering function of the yolks in a system
with no shell-shell interactions, $ u_{ss}(r)=0$, with the same
yolk-shell interactions $ u_{ys}(r)$ as before (see Eq.
(\ref{hys})), and with the yolk-yolk interactions given by
\begin{equation}
\beta u_{yy}(r)=\left\{ \begin{array}{ll}
\infty & \textrm{for $r<\sigma_{y}$} \\
K\frac{exp[-z(r/\sigma_s-1))]}{r/\sigma_s} & \textrm{for $r>
\sigma_{y}$},
\end{array} \right.
\label{hyyy}
\end{equation}
with $K=554$ and $z=0.149$.

\section{Summary}

In summary,  in this paper we systematically studied the
dependence of collective diffusion and self-diffusion properties
of the yolk-shell complex on the relative mobility (i.e., the
ratio $\zeta^0_s/\zeta^0_y$),  of the shell and yolk particles,
and on the concentration of yolk-shell complexes in the
suspension. For this, we assumed purely repulsive hard-body
interactions between all (shell and yolk) particles, and carried
out systematic Brownian dynamics (BD) simulations to determine the
pertinent self- and collective-diffusion properties. In addition,
we developed a first-principles theory to predict the main
features of the shell-shell dynamic properties (i.e., of $F(k,t)$
and $F_S(k,t)$, etc.) in terms of the specific effective
interactions between the shell and the  yolk particles that
constitute the system. We approached this task in the framework of
the self-consistent generalized Langevin equation (SCGLE) theory
of colloid dynamics, which was adapted here to the context of a
suspension of Brownian yolk-shell particles.

Our results,  as obtained from BD and as predicted by SCGLE
theory, show that the yolk-shell interaction affects the
properties of diffusion of the yolk-shell complex. In the dilute
case, our simulations showed that the deviation of the MSD of the
yolk-shell complex from the purely diffusive behavior of a empty
shell is an indication of the presence of a friction effect upon
displacement of the complex due to the yolk-shell interaction.
This effect upon the diffusion was more dramatic when we changed
the condition of the solvent between the shell and the yolk
$\delta$. For example, if the interior of the shell becomes more
viscous, so that the ratio $\delta$ decreases, then also the
diffusivity of the yolk-shell particle will decrease.  The
predictions of the SCGLE corroborates with the results in the
dilute case and follow with good agreement the MSD of our
simulations. In the case of the finite concentration, results describe not only how interactions
between yolk and shell influence the dynamic properties of the yolk-shell complex,
but also the how they are affected by interactions between the shells. We analyzed the
mean square displacement and self-intermediate scattering function
of a liquid of yolk-shell particles, and compared them with
properties of a liquid of empty shells.  We found that at low volume
fractions the difference between the dynamic of a yolk-shell
complex and the dynamic of an empty-shell, is determined only by
the yolk-shell friction, whereas at higher concentrations it is
dominated by the shell-shell interactions. Also, our theoretical approach predicted the same scenario
obtained for the simulation of the yolk-shell complex as for the
empty shell liquid, with a very good agreement.

\section{Acknowledgments}
This work was supported by  the U.S. Department of Energy, Office
of Basic Energy Sciences, Materials Sciences and Engineering
Division. This Research at the SNS at Oak Ridge National
Laboratory was sponsored by the Scientific User Facilities
Division, Office of Basic Energy Sciences, U.S. Department of
Energy.

\appendix

\section{Averaging over the yolk degrees of freedom.}

Let us consider the stochastic but fully microscopic dynamic
description of the correlated Brownian motion of the $N$ shell and
the corresponding $N$ yolk particles provided by Eqs. (\ref{eq1p})
and (\ref{eq1py}). The purpose of this appendix is to derive in
detail from this description the set of $N$ ``renormalized''
Langevin equations in Eq. (\ref{eq2ppp}), involving the positions
and velocities of only the shell particles. If one is interested
in observing only the motion of the shell particles, then in
principle one needs  to solve Eqs. (\ref{eq1py}) for the positions
and velocities of the yolk particles, and then substitute in the
Eqs. (\ref{eq1p}), averaging over the initial value of the yolks'
positions and velocities. This, however, can be done in an
alternative manner by re-writing Eq.\ (\ref{eq1p}) as
\begin{equation}
M_s{\frac{d{\bf v}_{i}(t)}{dt}} = -\zeta^0_{s}{\bf v}_{i}(t)+{\bf f}
^0_{i}(t)-\sum_{j\neq i}\nabla_i u_{ss}(|{\bf x}_{i}(t)-{\bf
x}_{j}(t)|) + \int d^3{\bf r} [\nabla
u_{ys}(r)]n_y^*(\textbf{r},t). \label{eq2}
\end{equation}
where $n_y^*(\textbf{r},t)$, defined as $n^*_y(\textbf{r},t)\equiv
n_y(\textbf{x}_i(t)+\textbf{r},t)$, with $n_y(\textbf{x},t)\equiv
\delta(\textbf{x}- \textbf{y}_i(t))$, is  the probability that at
time $t$ the center of the yolk particle is located at position
\textbf{r} (referred to the center of the $i$th shell particle).
This means that the direct  pairwise potential $u_{ys}(r)$
between the shell and its yolk couples {\it exactly} the motion of
the shell particle with the motion of its yolk  {\it only} through
the variable $n^*(\textbf{r},t)$.

Since Eq. (\ref{eq1py}) is an ordinary Langevin equation for a
Brownian particle in the confining field of the shell, it is
equivalent to a Fokker-Planck (FP) equation for the probability
$P^*(\textbf{v},\textbf{r};t)$ that the yolk has velocity
\textbf{v} and position \textbf{r} at time $t$. In the limit of
overdamping ($t\gg M^{(y)}/\zeta ^{(y)}$), and integrating out the
velocity of the yolk, such FP equation becomes the Smoluchowski
(or diffusion) equation for  a single Brownian particle in the
external potential  $u^{(ys)}(r)$, which can be written as the
following equation for $n^*_y(\textbf{r},t)= \int d^3v
P^*(\textbf{v},\textbf{r};t)$
\begin{equation}
{\frac{\partial  n^*_y(\textbf{r},t)}{dt}}=  [\nabla
n^*_y(\textbf{r},t)] \cdot {\bf v}_T(t) + D^0_y\nabla^2
n^*_y(\textbf{r},t) + D^0_y\nabla \cdot n^*_y(\textbf{r},t)\nabla
\beta u_{ys}(r) , \label{eqdify}
\end{equation}
where the first term on the \textit{r.h.s.} is a streaming term
(due to the fact that $n^*_y(\textbf{r},t)\equiv
n_y(\textbf{x}_T(t)+\textbf{r},t)$) and $D^0_y \equiv
k_BT/\zeta^0_{y}$ is the diffusion coefficient of the yolk
particle inside the shell tracer particle.

Eqs. (\ref{eq2}) and (\ref{eqdify}) are two coupled equations for
the variables ${\bf v}_T(t)$ and $n^*_y(\textbf{r},t)$, which
involve the non-linear streaming term $[\nabla
n^*_y(\textbf{r},t)] \cdot {\bf v}_T(t)$. What we need, however,
is the stochastic and linearized version (around the equilibrium
values ${\bf v}^{eq}_T=0$ and $n^{eq}_y(r)$) of Eqs.  (\ref{eq2})
and (\ref{eqdify}). For this, notice that $n^{eq}_y(r)$, is given
by
\begin{equation}
n_y^{eq}(r)=\frac{e^{-\beta u_{ys}(r)} }{\int  e^{-\beta
u_{ys}(r)} d^3r}, \label{neqy}
\end{equation}
and that, due to the radial symmetry of the force $[-\nabla
u_{ys}(r)]$ and of $n^{eq}_y(r)$, the second integral in Eq.\
(\ref{eq2}), evaluated at $n^*_y(\textbf{r},t)= n^{eq}_y(r)$,
vanishes. The resulting fluctuating linearized version of Eqs.
(\ref{eq2}) and (\ref{eqdify}) can thus be written as
\begin{equation}
M_s{\frac{d{\bf v}_{i}(t)}{dt}}= -\zeta^0_{s}{\bf v}_{i}(t)+{\bf f}
^0_{i}(t)-\sum_{j\neq i}\nabla_i u_{ss}(|{\bf x}_{i}(t)-{\bf x}_{j}(t)|)
+ \int d^3{\bf r} [\nabla u_{ys}(r)]\delta n_y^*(\textbf{r},t),
\label{eq2p}
\end{equation}
with $\delta n_y^*(\textbf{r},t)\equiv
n_y^*(\textbf{r},t)-n^{eq}_y(r)$, and
\begin{equation} {\frac{\partial \delta
n^*_y(\textbf{r},t)}{dt}}=  [\nabla n^{eq}_y(r)] \cdot {\bf
v}_T(t) + D^0_y\nabla^2 \delta n^*_y(\textbf{r},t) + D^0_y\nabla
\cdot \delta n^*_y(\textbf{r},t)\nabla \beta u_{ys}(r)  +
f^0_y(\textbf{r},t), \label{flucteqdify}
\end{equation}
where  $f^0_y(\textbf{r},t)$ is a Gaussian fluctuating term with
zero mean value and time-correlation function given by the
fluctuation-dissipation relation $\langle
f^0_y(\textbf{r},t)f^0_y(\textbf{r}',t')\rangle= [D^0_y\nabla^2 \delta
(\textbf{r}-\textbf{r}')]n^{eq}_y(r')2\delta(t-t')$.

Formally solving Eq.\ (\ref{flucteqdify}) and substituting the
solution for $\delta n^*_y(\textbf{r},t)$ in Eq.\ (\ref{eq2p}),
leads to
\begin{equation}
M_s{\frac{d{\bf v}_{i}(t)}{dt}}= -\zeta^0_{s}{\bf v}_{i}(t)- \int_0^t
dt' \stackrel{\leftrightarrow }{\Delta \zeta_y(t}-t') \cdot {\bf
v}_{i}(t')+{\bf F} _{i}(t)-\sum_{j\neq i}\nabla_i u_{ss}(|{\bf
x}_{i}(t)-{\bf x}_{j}(t)|), \label{eq2pp}
\end{equation}
where the new fluctuating force ${\bf F}_{i} (t)$ has zero mean
and correlation function given by the fluctuation-dissipation
relation $ \langle {\bf F}_{i}(t){\bf F}_{j}(0)\rangle
=k_{B}T[\zeta^0_{s}2\delta (t)+ \Delta \zeta _y(t) ]\delta
_{ij}\stackrel{\leftrightarrow }{{\bf I}}$ (with $i,j=1,2,\ldots
,N)$ and with the time-dependent friction function
$\stackrel{\leftrightarrow }{\Delta \zeta_y (t) }$ given by the
following \textit{exact} result:
\begin{equation}
\Delta \stackrel{\leftrightarrow }{\zeta}_y (t)= -\int d^3 r \int
d^3 r' [\nabla u_{ys}(r)] \chi _y
^*(\textbf{r}-\textbf{r}';t)[\nabla ' n^{eq}_y(r')]. \label{dzty}
\end{equation}
In this equation, $\chi_y ^* (\textbf{r}-\textbf{r}';t)$ is the
Green's function of the diffusion equation in Eq.\
(\ref{flucteqdify}), i.e., it is the solution of
\begin{equation} {\frac{\partial \chi_y ^* (\textbf{r}-\textbf{r}';t)}{dt}}
=   D^0_y\nabla^2 \chi_y ^* (\textbf{r}-\textbf{r}';t)+
D^0_y\nabla \cdot \chi_y ^* (\textbf{r}-\textbf{r}';t)\nabla \beta
u_{ys}(r) \label{flucteqdifygreens}
\end{equation}
with initial condition $\chi_y ^*
(\textbf{r}-\textbf{r}';0)=\delta (\textbf{r}-\textbf{r}')$.

Using the fact that $\nabla n^{eq}_y(r)= -\beta n^{eq}_y(r)\nabla
u_{ys}(r)$, the previous expression for $\Delta
\stackrel{\leftrightarrow }{\zeta}_y (t)$ can also be written as
\begin{equation}
\Delta \stackrel{\leftrightarrow }{\zeta}_y (t)= k_BT\int d^3 r
\int d^3 r' [\nabla n^{eq}_y(r)] \left\{ \frac{\chi _y
^*(\textbf{r}-\textbf{r}';t)}{ n^{eq}_y(r)}\right\}[\nabla '
n^{eq}_y(r')]. \label{dzdty1}
\end{equation}
This expression can be further simplified by introducing the
homogeneity approximation, which consists of approximating $[\chi
_y ^*(\textbf{r}-\textbf{r}';t)/ n^{eq}_y(r)]$ by $[\chi _y
^*(\mid\textbf{r}-\textbf{r}'\mid;t)/ n_0]$ in the integrand,
which allows us to introduce the Fourier transform (FT) $\chi _y
^*(k;t)$ of $\chi _y ^*(r;t)$, with the result
\begin{equation}
\Delta \stackrel{\leftrightarrow }{\zeta}_y (t)= \frac{k_BT
n_0}{(2\pi)^3}\int d^3 k  \textbf{kk}[g_y(k)]^2 \chi _y ^*(k;t)\
 . \label{dzdty200}
\end{equation}

At this point we introduce the so-called decoupling approximation,
explained in Ref. \cite{faraday}, which approximates $\chi _y
^*(k;t)$ by the product $\chi _y (k;t)F_S(k,t)$, in which
$F_S(k,t)$ is the shell intermediate scattering function and $\chi
_y (k;t)$ (without the asterisk indicating the moving reference
frame of the shell), which can be approximated by the Gaussian
approximation  with a simple short-time expression for the mean
squared displacement $W_y(t)$ of the yolk, namely, $\chi _y
(k;t)\approx \exp [-k^2D^0_y t]$. Using these approximations, and
taking into account that the tensor $\Delta
\stackrel{\leftrightarrow }{\zeta}_y (t)$ must be isotropic, so
that it can be written as $ \Delta \stackrel{\leftrightarrow
}{\zeta}_y (t)= \Delta \zeta_y (t)\stackrel{\leftrightarrow }{{\bf
I}}$, we may rewrite the yolk-averaged N-particle Langevin
equation in Eq. (\ref{eq2pp}) as
\begin{equation} M_s{\frac{d{\bf
v}_{i}(t)}{dt}}= -\zeta^0_{s}{\bf v}_{i}(t)- \int_0^t dt'\Delta
\zeta_y(t-t')  {\bf v}_{i}(t')+{\bf F} _{i}(t)-\sum_{j\neq
i}\nabla_i u_{ss}(|{\bf x}_{i}(t)-{\bf x}_{j}(t)|) \label{eq2ppp}
\end{equation}
with the scalar time-dependent friction function $\Delta \zeta_y
(t)$ given by one third of the trace of $\Delta
\stackrel{\leftrightarrow }{\zeta}_Y (t)$, i.e., by
\begin{equation}
\Delta\zeta_y(t)= \frac{k_BT n_0}{3(2\pi)^3}\int d^3 k [kg_y(k)]^2
e^{-k^2D^0_y t} F_S(k,t). \label{dzdty4}
\end{equation}

\end{document}